\documentclass[a4paper,11pt]{article}
\usepackage{jheppub}
\makeatletter
\def\@fpheader{Published in JHEP \textbf{11} (2026) 113, \href{https://doi.org/10.1007/JHEP11(2026)113
}{doi:10.1007/JHEP11(2026)113}}
\makeatother
\usepackage{lineno}
\usepackage{subcaption}
\usepackage{xcolor}
\usepackage{hyperref}
\usepackage{bm}
\usepackage{wrapfig,lipsum,booktabs}

\usepackage{tikz}
\usepackage{float}

\title{Gravitational Duals from Equations of State II: Large Hierarchies and False Vacua}

\author[b,c]{Raul Jimenez,}
\author[a,b,c]{David Mateos,}
\author[d]{Pavlos Protopapas,}
\author[a,b]{Pau Sol\'e-Vilar\'o,}
\author[a,b]{Pedro Taranc\'on-\'Alvarez,}
\author[a,b]{Pablo Tejerina-P\'erez}

\affiliation[a]{Departament de F\'\i sica Qu\`antica i Astrof\'\i sica,  Universitat de Barcelona, Mart\'\i\  i Franqu\`es 1, 
\mbox{ES-08028}, Barcelona, Spain.}
\affiliation[b]{Institut de Ci\`encies del Cosmos (ICC),  Universitat de Barcelona, Mart\'\i\  i Franqu\`es 1, 
\mbox{ES-08028}, Barcelona, Spain.}
\affiliation[c]{Instituci\'o Catalana de Recerca i Estudis Avan\c cats (ICREA), Passeig Llu\'\i s Companys 23, \\
\mbox{ES-08010}, Barcelona, Spain.}
\affiliation[d]{Institute for Applied Computational Science, Harvard University, 150 Western Avenue, Allston, MA 02134, U.S.A.}

\emailAdd{raul.jimenez@icc.ub.edu; dmateos@fqa.ub.edu; pprotopapas@g.harvard.edu; pau.sole@fqa.ub.edu; pedro.tarancon@fqa.ub.edu;  pablo.tejerina@icc.ub.edu}

\abstract{We investigate the reconstruction of holographic duals for strongly coupled quantum field theories in regimes characterized by large  hierarchies and the presence of false vacua.  
Within the gauge/gravity duality, these features translate into non-trivial thermodynamic behaviour and exotic renormalization group flows, including skipping flows between non-adjacent fixed points. Building on previous work based on Physics-Informed Neural Networks (PINNs), we extend the holographic inverse problem of reconstructing the bulk scalar potential from boundary thermodynamic data into this new regime. This setting presents a variety of conceptual and numerical challenges,  
such as near-degenerate states, large hierarchies of energy scales, and regions of the potential that are not directly probed by the input data. 
We develop a set of methodological advances that overcome these obstacles, thereby improving the established PINNs-based methodology and extending it to new physical regimes of interest that were previously out of reach. 
Applying the developed framework, we demonstrate accurate reconstruction of scalar potentials deep into the false vacuum regime, achieving robust agreement with the physical features of the underlying thermodynamics despite significant numerical stiffness. Our results extend the bridge between holography and machine learning, and suggest that data-driven approaches can provide new insights into the structure of 
strongly coupled systems.
}

\begin{document}
\maketitle
\flushbottom

\section{Introduction}
\label{sec: intro}

Holography \cite{Maldacena:1997re,Gubser:1998bc,Witten:1998qj,aharony2000large} maps the quantum properties of strongly coupled gauge theories with a large number of colours to the classical properties of gravitational theories in one higher dimension. Under this map, the thermodynamic properties of the gauge theory are encoded in properties of black hole horizons in the dual gravitational solutions. In particular, the equation of state of the field theory, encoded in the thermodynamic relation between entropy density and temperature, $S(T)$, is determined by the set of all static black hole solutions. Nevertheless, the most far-reaching consequence of the duality is arguably that even arbitrarily-far-from-equilibrium dynamics of the gauge theory can be determined by evolving the dual Einstein's equations in time. In such dynamical regimes, holography often provides the only available framework for controlled first-principle calculations, in contrast to equilibrium settings where alternative approaches, such as lattice gauge theory, can sometimes be employed.

Hence, an appealing strategy is to construct a holographic model that reproduces the equilibrium properties of a gauge theory of interest, and then use the gravitational dual to access the out-of-equilibrium dynamics. 
However, implementing this strategy requires solving a challenging inverse problem: given the equilibrium equation of state of the gauge theory one must determine the dual gravitational theory.

Early (semi)analytical attempts to this inversion problem include \cite{Gubser:2008ny,Gubser:2008yx,Alanen:2009xs} (see \cite{Rougemont:2023gfz} for a review). In a previous paper \cite{Bea:2024xgv}, we presented a novel approach to solve this inverse problem based on Physics-Informed Neural Networks (PINNs), focusing on the case where the gravitational theory is an Einstein-Klein-Gordon (EKG) model in five dimensions. Our algorithm reconstructed not just a specific black hole solution but the whole gravitational theory itself, encoded in the scalar potential $V(\phi)$ of the EKG model. This was successfully demonstrated for a family of four-dimensional Conformal Field Theories (CFTs) deformed by a relevant operator of conformal dimension $\Delta = 3$, whose dual potential was parametrised by a single parameter $\phi_M$ that hence controlled the nature of the CFT thermal phase transition. The method was applied to theories exhibiting crossovers ($\phi_M = 5$), second-order ($\phi_M = 1.08$) and first-order ($\phi_M = 1$) phase transitions, achieving sub-percent relative errors in the recovered potential. In \cite{tarancon2025efficient}, the framework was further developed through the introduction of novel features in the neural network (NN) pipeline, such as multi-head training and unimodular regularisation techniques, that improved the efficiency of the transfer learning process across different parameter regimes.

However, the approaches used in both aforementioned works proved insufficient to accurately recover potentials with values of $\phi_M \lesssim 0.8$. As was shown in \cite{bea2018heating}, for values of $\phi_M$ below a critical threshold $\phi_M \simeq 0.5808$, due to its functional form the scalar potential develops additional features in the form of two new local extrema between its original global maximum and minimum. 

\begin{figure}[t]
    \centering
    \begin{subfigure}[b]{0.49\textwidth}
        \centering
        \includegraphics[width=\textwidth]{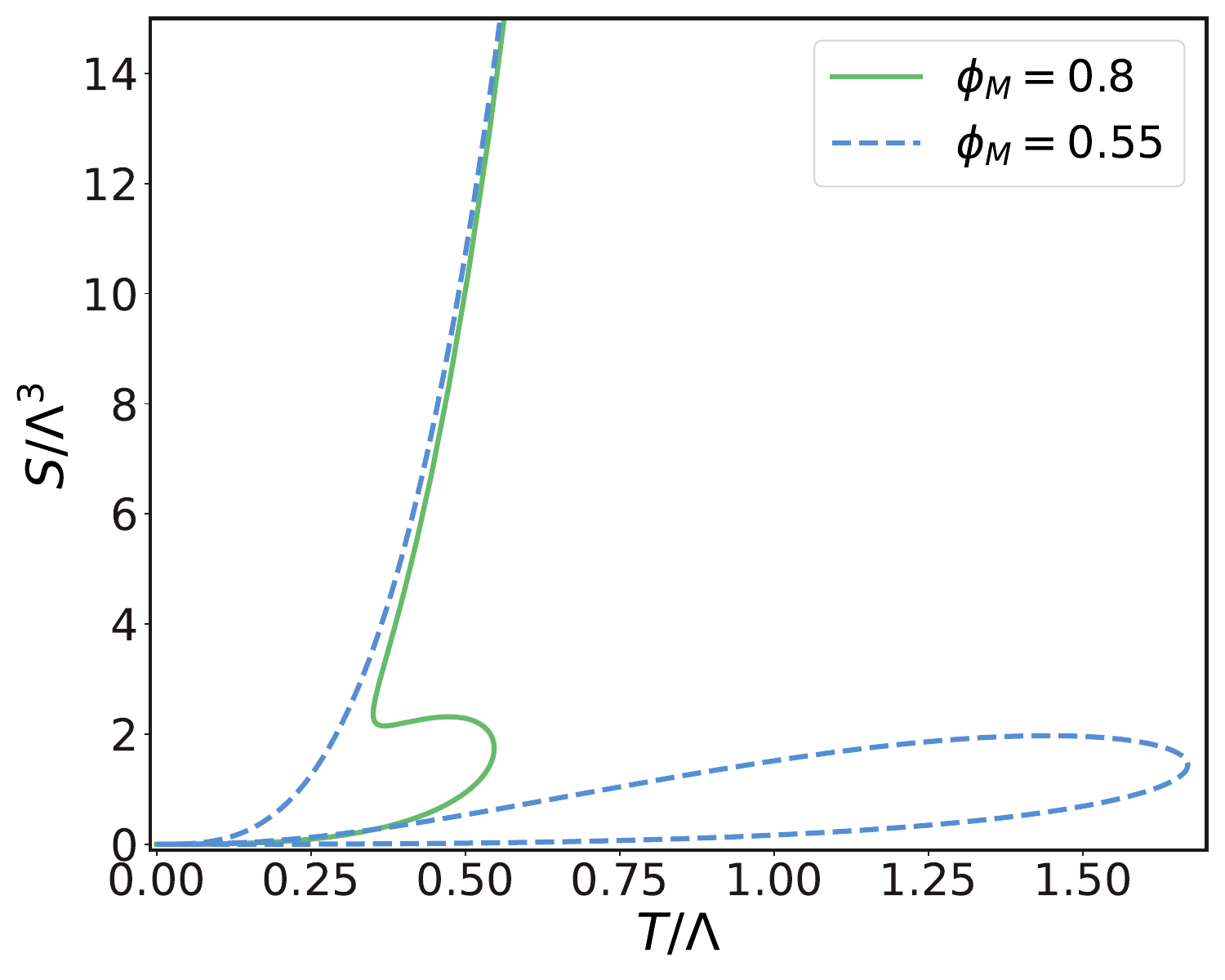}
        \caption{\small }
        \label{subfig:SofT}
    \end{subfigure}
    \hfill
    \begin{subfigure}[b]{0.49\textwidth}
        \centering
        \includegraphics[width=\textwidth]{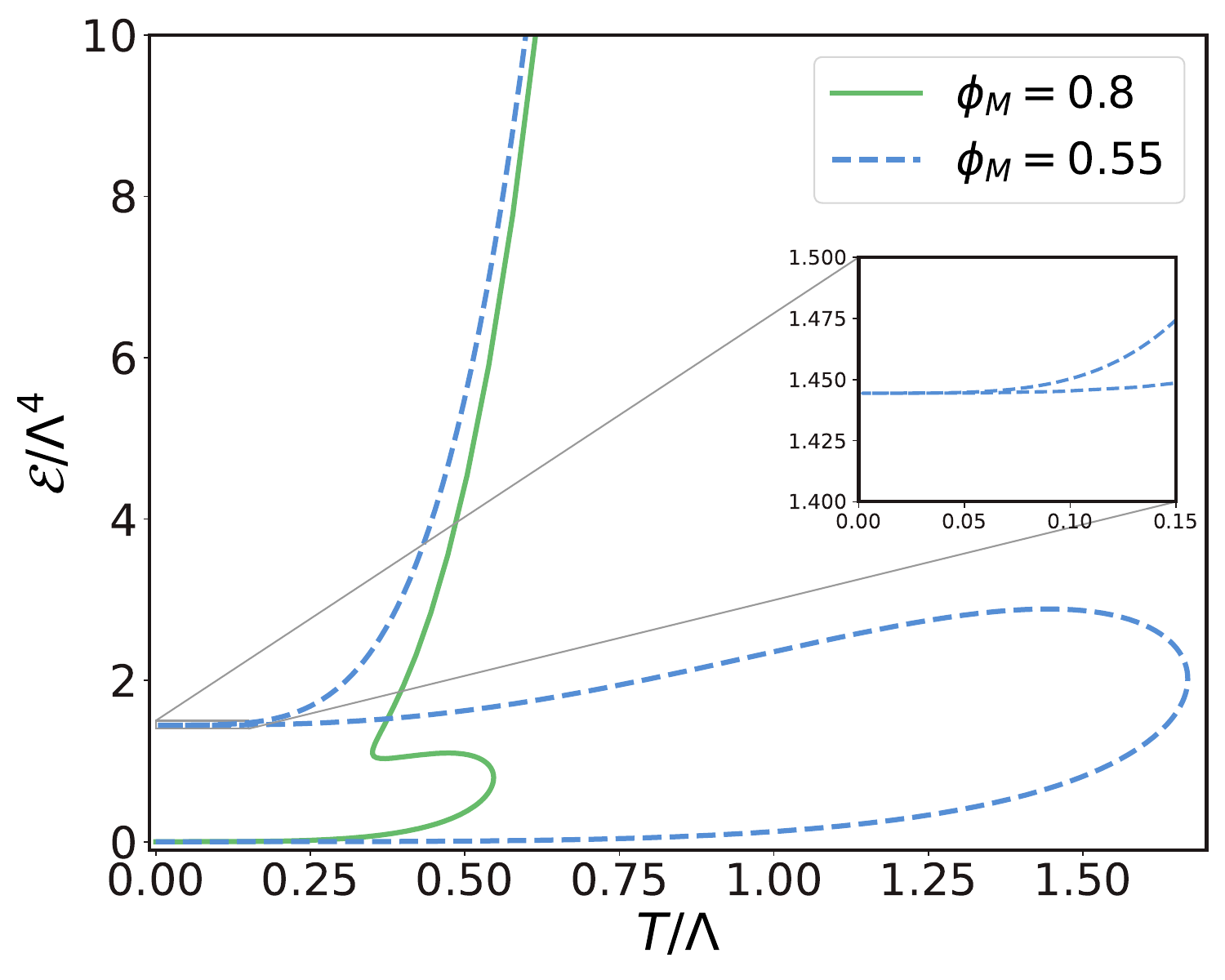}
        \caption{\small }
        \label{subfig:EofT}
    \end{subfigure}\\
    \centering
    \begin{subfigure}[c]{0.49\textwidth}
        \centering
        \includegraphics[width=\textwidth]{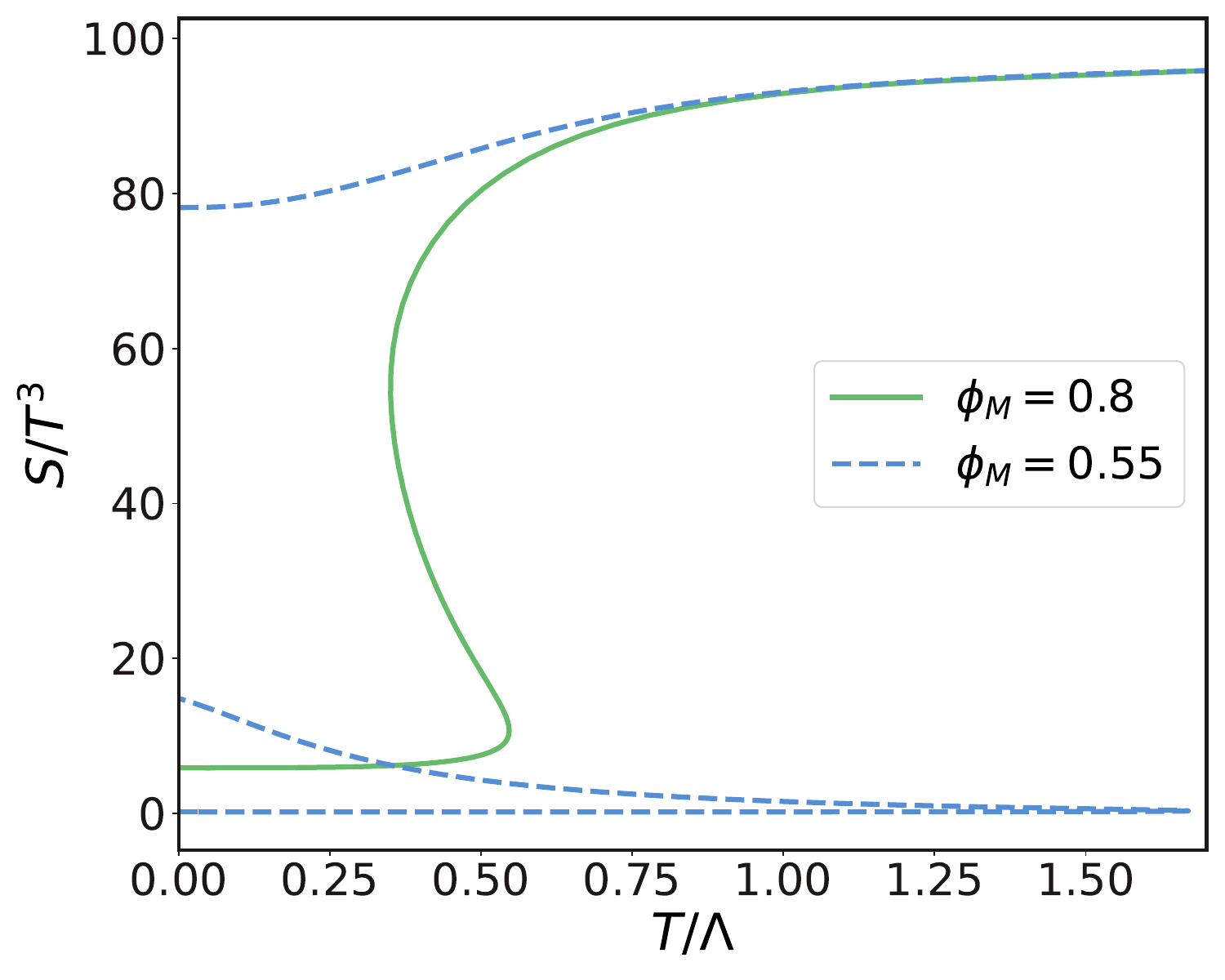}
        \caption{\small }
        \label{subfig:soverT3}
    \end{subfigure}
    \begin{subfigure}[d]{0.49\textwidth}
        \centering
        \includegraphics[width=\textwidth]{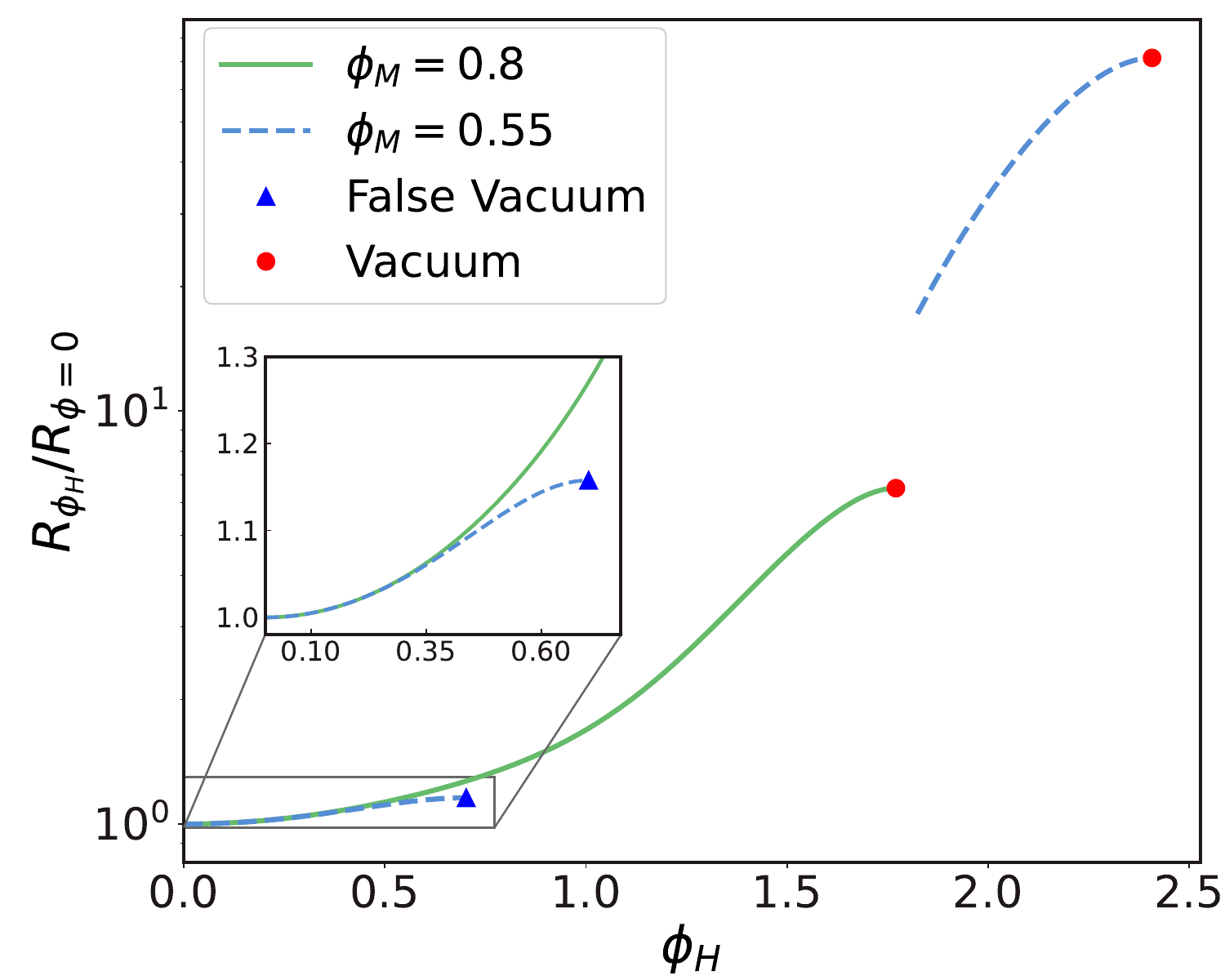}
        \caption{\small }
        \label{subfig:Ricci}
    \end{subfigure}
    \caption{\small Entropy density (a), energy density (b), and entropy-to-temperature cubed ratio (c) as a function of temperature for two different values of the parameter $\phi_M$. The value $\phi_M = 0.8$ corresponds to a first-order phase transition. As we decrease the value of this parameter, the transition becomes more pronounced and develops a false vacuum state, as seen for the case of $\phi_M = 0.55$. At $T/\Lambda = 0$ there are two available states: the false vacuum and the true vacuum states, as seen in the dashed blue lines in (b). In (d) we plot the value of the Ricci scalar in the bulk, evaluated at the horizon and normalized by its asymptotic AdS value, as a function of the scalar field at the horizon. The red and blue dots represent the true and false vacua, respectively. The gap in the blue dashed curve reflects the absence of solutions in that region. In both theories, a large separation of scales between their respective vacua can also be observed.}
    \label{fig:EofT}
\end{figure}

This has important implications across various aspects in the problem at hand. On the physics side, potentials below this critical threshold are dual to boundary theories that possess a \textit{false vacuum}, namely a zero-temperature, Lorentz-invariant state with energy density higher than that of the true vacuum. A comparison between such a boundary theory and another one without a false vacuum can be seen in Fig.~\ref{fig:EofT}. Without loss of generality, we will assign zero energy to the true vacuum, so the energy of the false vacuum will be positive. In the holographic context this can be achieved by an appropriate choice of counterterms. The gravitational theories dual to these QFTs can give rise to so-called exotic renormalization group (RG) flows when looking at the behaviour of the running coupling at the boundary \cite{bea2018heating,kiritsis2017exotic}. For instance, we can find ``skipping flows'' that connect non-adjacent fixed points while bypassing intermediate CFTs, which is unusual in field theory, where flows stop at the first infrared (IR) fixed  point available. Finally, the thermodynamics at the boundary become remarkably non-trivial, with the equation of state developing a complex multi-valued structure that encodes information about all the intermediate CFTs in the potential landscape.

On the computational side, the false vacuum regime presents qualitatively new challenges that go beyond mere numerical stiffness. As will be explained in further detail below, the scalar potentials in this regime will now contain a region whose dual thermal states in fact belong to a different boundary theory and are therefore absent from the equation of state given as an input to the PINNs model, as seen pictorially in 
Fig.~\ref{subfig:Ricci}. Additionally, the false and true vacuum states correspond to the same point (the origin)  in the $(S, T)$ plane (see Fig.~\ref{subfig:SofT}) despite corresponding to very different bulk solutions, creating a degeneracy that must be resolved. Finally, there is a large separation of scales between the defining features of the potential; the local extrema are shallow compared to the deep global minimum, and hence represent small-scale features present in the solution (see Figs.~\ref{fig:Potentials} and \ref{subfig:Ricci}). This issue in particular is well-known in the differential equation (DE) mathematical and numerical literature \cite{hairer1996solving, Seiler2025}, and in our context is prevalent even in simpler cases outside of the false vacuum regime. It is presumably one of the main reasons why the PINNs-based strategy used in previous works had so far not yielded satisfactory results when applied to the near-false-vacuum regime of $0.5808 \leq \phi_M \leq 0.8$, where one might hope it could still be applicable.

Some of the challenges discussed above stem in part from the fact that the scalar potential used to benchmark our algorithm is derived from a globally well-defined superpotential. Relaxing this assumption could mitigate certain features, such as the pronounced hierarchy between the false and true vacua. However, rather than simplifying the setup, we choose to treat these difficulties as a virtue, using them to probe the limits of the reconstruction framework. Building on the previous successes achieved using PINNs in this holographic inverse problem \cite{Bea:2024xgv, tarancon2025efficient}, a natural next step is to extend the methodology to access these more intricate and physically rich regimes. In this work, we address the challenges outlined above through a series of targeted and notable improvements to the PINNs-based algorithm.

Other applications of machine learning (ML) to holography include the use of NNs as representations of holographic renormalization and emergent geometry \cite{Gan:2017nyt,Hashimoto:2019bih,Hu:2019nea}, as well as data-driven reconstruction of particular bulk geometries or specific gravitational \textit{solutions} from boundary observables \cite{You:2017guh,Hashimoto:2018ftp,Hashimoto:2018bnb,Han:2019wue,Tan:2019czc,Yan:2020wcd,Akutagawa:2020yeo,Hashimoto:2020jug,Jokela:2020auu,Song:2020agw,Hashimoto:2021umd,Li:2022zjc,Jokela:2022fvh,Park_2022,Jokela:2023yun,Xu:2023eof,Park:2023slm,Ahn:2024gjf,Gu:2024lrz,jejjala2025learningholographichorizons,kim2025learninginverseryutakayanagiformula}. By contrast, the PINNs-based algorithm introduced in \cite{Bea:2024xgv,tarancon2025efficient} and the extensions developed in this work aim at reconstructing the holographic gravitational \textit{theory} itself within a prescribed field content. Related inverse-theory approaches based on differentiable forward solvers, rather than PINNs, have been developed in \cite{Hashimoto:2021ihd,Hashimoto:2022eij,Chen:2024ckb}.

The rest of the work is organised as follows. In Section \ref{sec 2: holomodel}, we review the holographic model including the functional form of the potential used and the formulation of the direct and inverse problems, with an emphasis on the new theoretical and technical aspects that arise in the false vacuum regime. The methodology used is presented in Section \ref{sec 3: methodology}, including the architectures and novel techniques employed, as well as the training procedures. In Section \ref{sec 4: results}, we present the results of the potential reconstruction for two cases in the regimes of interest, namely $\phi_M = 0.8$ and $\phi_M = 0.55$. Finally, results and an outlook on future directions are discussed in Section \ref{sec 5: discussion}. Our full code is publicly available in the following  GitHub repository: \href{https://github.com/pedrota2000/NNHolo}{NNHolo} \cite{NNHolo}.

\section{Holographic model}
\label{sec 2: holomodel}

\subsection{The potential}
\label{subsec: setup}
As outlined in the previous section, in this work we continue to explore the bottom-up gravity model which had been successfully recovered, within a certain parameter regime, from the thermodynamics of the dual boundary theory in our earlier work \cite{Bea:2024xgv}. The action of the five-dimensional EKG model takes the form
\begin{equation}
    S=\frac{2}{\kappa_5^2} \int dx^5\sqrt{-g}\left[\frac{1}{4} R -\frac{1}{2}(\nabla \phi)^2 - V(\phi) \right] \, ,
    \label{EinsteinScalarAction}
\end{equation}
where $\kappa_5$ is the gravitational coupling, which hereafter we will set to $\kappa_5^2=2$. $V(\phi)$ is the scalar potential, derived from a so-called superpotential $W(\phi)$ such that
\begin{equation}
    V(\phi) = -\frac{4}{3}W(\phi)^2 +\frac{1}{2}W'(\phi)^2.
\end{equation}

\begin{figure}
    \centering
    \begin{subfigure}[b]{0.49\textwidth}
        \centering
        \includegraphics[width=\textwidth]{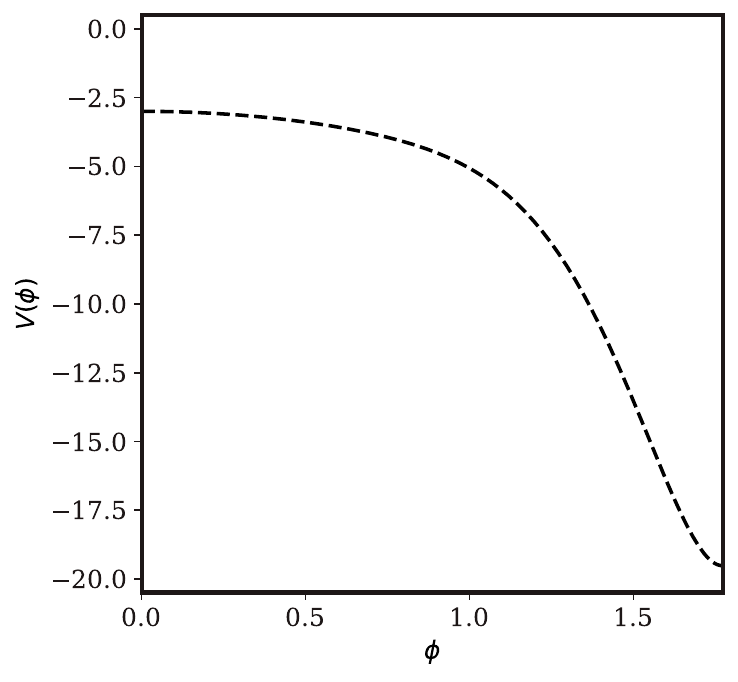}
        \caption{\small }
        \label{fig:potential1}
    \end{subfigure}
    \hfill
    \begin{subfigure}[b]{0.49\textwidth}
        \centering
        \includegraphics[width=\textwidth]{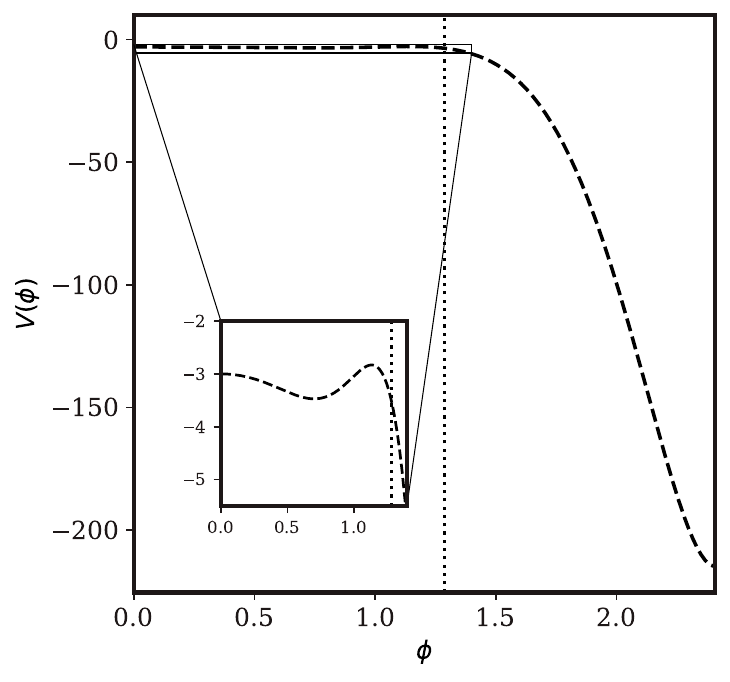}
        \caption{\small }
        \label{fig:potential2}
    \end{subfigure}
    \caption{\small The form of the potential $V(\phi;\phi_M)$ for different choices of the parameter $\phi_M$. On the left, the potential for $\phi_M = 0.8$ corresponds to a QFT at the boundary with a $1^{st}$-order phase transition (a). On the right, we enter the false vacuum regime with $\phi_M = 0.55$ as the potential develops an additional local minimum and maximum in the zoomed-in region (b). Notice the large separation of scales between features of the potential in (b), as well as the order of magnitude increase in the range of the potential when compared to (a). The parameter $\phi_c$ is found numerically and will be introduced in Section \ref{sec: direct problem}.}
    \label{fig:Potentials}
\end{figure}

Here, it is important to note that although any extremum of $W(\phi)$ is necessarily an extremum of $V(\phi)$, the reverse does not generally hold. In particular, extrema of both the potential and superpotential will be dual to supersymmetric CFTs, while extrema of the potential but not of the superpotential will be dual to non-supersymmetric 
CFTs.\footnote{Assuming that our bosonic theory is the truncation of a supersymmetric theory.} Following \cite{bea2018heating}, we choose the superpotential
\begin{equation}
    W(\phi) = -\frac{3}{2} - \frac{\phi^2}{2} - \frac{\phi^4}{4\phi^2_M} + \frac{\phi^6}{10},
    \label{superpotential}
\end{equation}
where we have set the asymptotic AdS radius to unity. This is a particular choice within the formalism at hand, which has been used in a variety of works to connect Einstein's equations with the holographic RG \cite{de2000holographic, kiritsis2017exotic, kiritsis2014holographic}. The value of  the parameter $\phi_M$ determines the form of the potential and hence the properties of the dual boundary theory, such as its RG flow or its thermal structure. For $\phi_M > 1.08$, the thermal physics exhibits a crossover between two conformal regimes controlled by an ultraviolet (UV) and an IR fixed point; as we lower the parameter, this becomes a $2^{nd}$-order phase transition at $\phi_M \simeq 1.08$, and then a $1^{st}$-order phase transition for $\phi_M \lesssim 1.08$. For these cases, the potential has the usual form involving one maximum and one minimum (in the region of interest), both of which   are also extrema of the superpotential (see Fig.~\ref{fig:potential1}), each corresponding to AdS solutions of different radii.

The phase transition becomes more pronounced as $\phi_M$ decreases further, entering the near-false-vacuum regime below $\phi_M = 0.8$. Eventually we reach a point at \mbox{$\phi_M \simeq 0.5808$} where the potential develops a degenerate critical point at $\phi \simeq 0.9015$, at which both the first and second derivatives of $V(\phi)$ vanish. Beyond this value, we enter the false vacuum regime where the potential develops two additional features in the form of a local minimum and maximum between the two original extrema. These new critical points are extrema of the potential but not of the superpotential. The true vacuum of the gauge theory is associated to an RG flow in the bulk that starts at the original maximum and ends at the original minimum. The false vacuum is associated to a flow from the original maximum to the new local minimum. The general form of these two possible morphologies of the potential, with the global and local stationary points, is shown in Fig.~\ref{fig:Potentials}.

\subsection{The direct problem}
\label{sec: direct problem}

In the context of the gauge/gravity duality, the scalar potential $V(\phi)$ in the bulk action \eqref{EinsteinScalarAction} encodes all characteristics of the gauge theory, including its thermodynamic behaviour. To extract the thermodynamics from a given potential requires solving the so-called direct problem, which involves finding every static, smooth black hole configuration that satisfies the EKG field equations. To this end, we take the following ansatz for the bulk metric in Eddington-Finkelstein coordinates:

\begin{equation}
ds^2=-A dt^2+\Sigma^2(dx^2+dy^2+dz^2)-\frac{2}{u^2}\, dt du \,\, ,
\label{metric_ansatz}
\end{equation}
where $\{t,x,y,z\}$ are boundary coordinates and $u$ is the null holographic coordinate such that $A$, $\Sigma$, and the bulk scalar field $\phi$ are functions of $u$. The AdS boundary is located at $u=0$. Using this ansatz to solve the bulk equations of motion derived from \eqref{EinsteinScalarAction}, we obtain the following ordinary differential equations (ODEs):

\begin{subequations}
\label{Einstein-scalar_equations}
\begin{eqnarray}
 \Sigma^{\prime \prime}+\frac{2}{u} \Sigma^{\prime}+\frac{2}{3} \Sigma \phi^{\prime 2}  &= & 0  \,\, , \\[2mm]
A^{\prime \prime}+\frac{8}{3 u^4} V(\phi)+A^{\prime}\left(\frac{2}{u}+\frac{3 \Sigma^{\prime}}{\Sigma}\right)  &= & 0  \,\, , \\[2mm]
\phi^{\prime \prime}-\frac{1}{u^4 A} \frac{\partial V(\phi)}{\partial \phi}+\phi^{\prime}\left(\frac{2}{u}+\frac{A'}{A}+3 \frac{\Sigma^{\prime}}{\Sigma}\right) & = &0  \,\, , \\[2mm]
 A^{\prime}+2 A \frac{\Sigma^{\prime}}{\Sigma}-\frac{2 \Sigma}{3 \Sigma^{\prime}}\left(A \phi^{\prime 2}-\frac{2}{u^4} V(\phi)\right) & = & 0 \,\, ,
 \label{constraint}
\end{eqnarray}
\end{subequations}
where a prime indicates differentiation  with respect to $u$.

To solve the equations of motion we must fix the sole input in the equations, namely the potential $V(\phi)$, and impose appropriate boundary conditions (BCs) that ensure that near the boundary at $u=0$ the geometry is asymptotically AdS. Hence, the near-boundary behaviour of the bulk functions is:
\begin{subequations}
\label{UVexpansion}
\begin{eqnarray}
 A(u)  &= & \frac{1}{u^2}+ \cdots   \,\, , \label{UVexpansion_a}\\[2mm]
\Sigma(u) &= & \frac{1}{u}+ \cdots  \,\, , \label{UVexpansion_b}\\[2mm]
\phi(u) & = & \Lambda u + \cdots  \,\,. \label{UVexpansion_c} 
\end{eqnarray}
\end{subequations} 
We have set the coefficients of the leading terms in $A$ and $\Sigma$ to unity without loss of generality, and  the dots indicate subleading terms in the limit $u\to 0$. The boundary condition for the scalar field $\phi$ imposes the fact that the CFT is deformed by a relevant  operator $\mathcal{O}$ with dimension $\Delta = 3$ and source $\Lambda$. The latter is the only intrinsic scale in the resulting QFT, and we will set $\Lambda=1$ hereafter.

For any value of the scalar field at the horizon, $\phi(u=u_H) \equiv \phi_H$, imposing these asymptotic functional forms leads to a unique pair $\{A_1, \Sigma_0\}$ of leading-order coefficients related to the surface gravity and the area density of the black brane horizon, from which the temperature $T$ and entropy density $S$ can be readily calculated:
\begin{subequations}
\label{temperature_entropy}
\begin{eqnarray}
 T  &= & \frac{A_1}{4 \pi}    \,\, , \label{temperature_entropy_a}\\[2mm]
S &= & \pi \Sigma_0^3  \,\,.
\label{temperature_entropy_b}
\end{eqnarray}
\end{subequations}
Thus, we can construct the thermodynamic curve $S(T)$ associated to thermal states at the boundary by numerically integrating the EKG equations \eqref{Einstein-scalar_equations} from the horizon to the boundary for each value of $\phi_H$, such that this parameter takes on the role of indexing the set of solutions. This \textit{direct} problem can be easily carried out once $V(\phi;\phi_M)$ has been specified, highlighting the key role of the potential in the setup. It is important to note that although in theories where a phase transition is present there will be multiple states $(S,T)$ with the same $T$, it still remains true that each will be uniquely associated to a value of $\phi_H$. 

However, as we enter the false vacuum regime by lowering the value of the free parameter to $\phi_M < 0.5808$, we begin to depart from this standard picture. Indeed, these potentials will now contain a range of values of $\phi_H$ whose solutions do not correspond to thermal states of the original QFT defined by the maximum at $\phi=0$, but to thermal states of the QFT defined by the new local maximum in the potential. The  values of $\phi_H$ that do not map to thermal states of the original QFT range from the new local minimum at $\phi = \phi_{FV}$ to some critical point $\phi = \phi_c$ that lies past the new maximum, as illustrated in Fig.~\ref{fig:potential2}. Nevertheless, the form of the potential within this range of values of the scalar field still affects some solutions that do correspond to thermal states of the original QFT. The reason is as follows.  From $\phi_c$ until the global minimum of the potential, the solutions to the direct problem once again correspond to thermal states of the $S(T)$ curve. To obtain these solutions one must integrate the EKG equations 
from $\phi_H>\phi_c$ back to $\phi=0$. Since this integration region  includes the region $(\phi_{FV}, \phi_c)$, this part of the $S(T)$ curve is sensitive to the value of the potential in the range of values of $\phi_H$ that does  not give rise to thermal states of the original QFT. In this theory, solutions with $\phi_H$ corresponding to either of the two minima or to $\phi_c$ all correspond to  states with zero entropy and temperature. 
The degenerate nature of these points, along with the portion of the potential that corresponds to thermal state solutions of a different theory, significantly increases the difficulty of solving the direct problem numerically by integrating the equations.

\subsection{The inverse problem}
\label{sec: inverse problem}
The inverse problem consists of recovering the bulk theory given only the boundary data; that is, recovering the scalar potential $V(\phi)$ characterising the action \eqref{EinsteinScalarAction} that is dual to a known equation of state $S(T)$. As we have seen, each pair of values $(T,S)$ in the thermodynamic curve represents a thermal state making up the boundary data of the QFT, and each state is dual to a planar black hole solution in the bulk with the same entropy density and temperature. Then, as done in the direct problem, a given pair of values for $(T,S)$ sets boundary values (at $u=u_H=1$) for the metric functions $A(u)$ and $\Sigma(u)$ in \eqref{metric_ansatz}. The relation is given in Eqs. (\ref{temperature_entropy_a}, \ref{temperature_entropy_b}).

Formulating the problem in this way, the potential $V(\phi)$ is an unknown, free function of the scalar field $\phi$, which in turn is one of the three functions that appear in the EKG equations \eqref{Einstein-scalar_equations}. The challenge now becomes inverting this free function $V(\phi)$, or in other words, finding $V(\phi)$ such that the ODEs admit solutions for all the BCs set by the different $(S,T)$ values. Evidently, this inversion of $V(\phi)$ has to be done in parallel to finding solutions for $\Sigma(u),A(u)$ and $\phi(u)$ that fulfill each of the different BCs. As one might expect, there will be as many solutions as BC instances (or equivalently, points in which the $S(T)$ curve is sampled), but all of them need to share the same free function $V(\phi)$ in the ODEs. This strong requirement, along with the fact that the ODEs at hand are coupled and highly non-linear, are in part what makes the inverse problem so numerically challenging, and what motivated the adoption of a PINNs framework to tackle it in earlier works. Moreover, this impracticability of using traditional numerical methods is especially true for recovering potentials in the false vacuum regime, where part of said potential will not be directly encoded in the initial data; as a result, in those cases any form of iterative numerical approaches that rely on prior solutions, such as certain applications of spectral or relaxation methods, will be effectively rendered out of use despite their general suitability and accuracy in solving other equally difficult numerical problems.

In principle, there is no mathematical guarantee that there exists such a function, or that there are no degeneracies related to it. However, one may expect a unique solution to exist for physically-motivated $S(T)$'s, and this is plausible mathematically since we wish to constrain one function by specifying another. To simplify the problem, in this paper we will make the mild assumption that $S(T) \approx T^3$ both at high and low temperatures, as corresponds to a QFT with an RG flow between a UV and an IR fixed point.

\section{Methodology}
\label{sec 3: methodology}
In this section, we outline the methodology employed to reconstruct the potential and solve the EKG equations using PINNs. We begin by reviewing the relevant background, followed by introducing the specific equations to be solved and describe their formulation within this framework. We then discuss the ML implementation of the problem in the context of PINNs. Finally, we highlight several technical modifications introduced in our approach which depart from the standard PINN setup.

\subsection{Background}
PINNs were first introduced in the pioneering works of Dissanayake and Phan-Thien \cite{dissanayake_neural-network-based_1994} and subsequently refined by Lagaris et 
al.~\cite{lagaris_artificial_1998,lagaris_neural-network_2000}. Since then, PINNs have been established as a versatile framework that can be used to solve partial and ordinary differential equations (PDEs/ODEs). Their efficacy in addressing a broad spectrum of challenging physical problems was rigorously demonstrated by Raissi et al.~\cite{raissi2019physics}, while further methodological and applied advancements have been reported in \cite{mattheakis2021hamiltonian,sirignano2018dgm,zhu2019physics}. Within this paradigm, each unknown function in the governing equations is typically represented by a dedicated neural network, most commonly implemented as a multilayer perceptron (MLP) consisting of stacked, fully-connected layers interleaved with non-linear activation functions. The training procedure consists of minimizing a loss functional constructed from the squared residuals of the differential equations, thereby embedding the governing physical laws directly into the optimisation objective. For alternative strategies used to incorporate physical constraints into ML models, the reader is referred to \cite{Choudhary2019,Choudhary2020,Greydanus2019}.

Although PINNs still generally rank below traditional numerical methods when it comes to computational efficiency and accuracy, they present novel benefits. For instance, they offer closed-form approximations to solutions, thereby avoiding the reliance on traditional iterative solvers and significantly reducing computational costs. They are mesh-free, meaning that they enable on-demand evaluation of solutions once training is complete, which is particularly advantageous in the treatment of complex systems. Moreover, their capacity to exploit transfer learning (TL) facilitates the rapid identification of new solutions by leveraging knowledge acquired from related previous tasks, thus improving adaptability and accelerating convergence. An additional strength of PINNs lies in their invertibility, making them well-suited for inverse problems where reconstructing inputs from observed outputs is required. Furthermore, the framework can be extended to include equation parameters as inputs, allowing for a unified treatment of both state variables and system parameters within the same architecture.

One crucial element of our setup are the so called ``solution bundles''. This technique was first developed in \cite{flamant2020solving} as an extension of Lagaris' original method \cite{lagaris_neural-network_2000}. This consists of training the neural network (NN) on a variety of solutions at once, usually parametrized by different initial/boundary conditions or constants appearing on the equations themselves. As a result, the trained network can be reused more effectively, enabling faster execution of tasks that involve evaluating numerous solutions, including Bayesian parameter inference, uncertainty propagation in dynamical systems, and various classes of inverse problems.

\subsection{Einstein-Klein-Gordon equations}
We will now write the EKG equations \eqref{Einstein-scalar_equations} in a more suitable way for the PINNs pipeline. First, we will redefine the unknown functions $A(u)$, $\Sigma(u)$, $\phi(u)$ such that they are bounded in the entire computational domain $u\in\left[0,1\right]$. We can remove the singular behaviour near $u = 0$ seen in expression  \eqref{UVexpansion}
by introducing the following field redefinitions:
\begin{equation}
\tilde{\Sigma} = u \, \Sigma \,\, ,
 ~~~~   
\tilde{A} = u^2 A  \, .
\label{RedefinitionsASigma}
\end{equation}

From a computational point of view, it is more efficient for the PINNs setup to deal with first-order ODEs at most. To this end, the system of equations \eqref{Einstein-scalar_equations} can be rewritten by introducing $\tilde{A}, \tilde{\Sigma}$ and $\phi$ as independent variables. Together with the constraint \eqref{constraint}, this results in six functions $\{ \tilde{\Sigma}(u), \tilde{A}(u), \phi(u), \nu_{\Sigma}(u), \nu_A(u), \nu_{\phi}(u)  \} \equiv \vec{\psi}(u) $ subject to the following seven coupled first-order ODEs 
\begin{equation}
    E_\alpha=0 \,,\qquad \alpha=1, \ldots, 7\,,
    \label{Ealpha}
\end{equation}
where
\begin{subequations}
\label{1s_order_ODEs}
\begin{eqnarray}
E_1 & = & \nu_{\Sigma} - \tilde{\Sigma}'   \,,  \\[2mm]
E_2 & = & \nu_A - \tilde{A}' \,, \\[2mm]
E_3 & = & \nu_\phi - \phi'  \,, \\[2mm]
E_4 & = & \nu_{\Sigma}' +  \frac{2}{3} \tilde{\Sigma} \, 
 \nu_{\phi}^2  \,,\\[2mm]
E_5 & = & u^2 \, \tilde{\Sigma} \, \nu_A^{\prime} + \frac{8}{3} \, V(\phi)\, \tilde{\Sigma}+ 
    \nu_A \, \left(  3u^2 \, \nu_{\Sigma} - 5u\, \tilde{\Sigma}  \right) + 
           \tilde{A} \left( 8\tilde{\Sigma} - 6u \,\nu_{\Sigma} \right) \,, \\[2mm] 
E_6 & = & \frac{u^2 \,\tilde{\Sigma} \, \tilde{A}\, \nu_\phi'}{\text{max}\left(\nu_\phi,10^{-3}\right)} - \frac{\tilde{\Sigma}}{\text{max}\left(\nu_\phi,10^{-3}\right)}\,  \frac{d V}{d \phi} 
        +  \left( -3 u \, \tilde{A} \, \tilde{\Sigma}  + u^2 \,\tilde{\Sigma} \,\nu_A 
        + 3 u^2\, \nu_{\Sigma} \, \tilde{A} \right)  \,,  \\[2mm]     
E_7 & = & \left( u \, \nu_{\Sigma}-\tilde{\Sigma} \right) 
\left( u^2 \, \tilde{\Sigma} \, \nu_A + 2 u^2  \, \tilde{A} \, \nu_{\Sigma} - 4 u \tilde{A} \tilde{\Sigma} \right)
- \frac{2}{3}  u \, \tilde{\Sigma}^2 
\left( u^2 \tilde{A} \, \nu_\phi^2 - 2 V(\phi) \right) \,. \,\,\,\,\,\,       
\end{eqnarray}
\end{subequations}
Here we can indeed observe the free function $V(\phi)$ along with its first derivative, such that for any given $V(\phi)$ one would obtain different solutions $\vec{\psi}(u)$. Note that $E_6$ is the only equation in which there are terms being divided. This was done in order to avoid the approximate trivial solution to which the PINNs setup was observed to converge to during training tests, which was to set $\nu_\phi = 0$. In particular, the model was finding that it was beneficial in terms of reducing the loss function to set this function to zero, even if this solution was wrong, as this then reduced $E_6$ to a much simpler equation to solve. The choice between $\nu_\phi$ and $10^{-3}$ ensures that the division does not blow up if the model tries to set a low value for $\nu_\phi$.

The BCs discussed in Sec.~\ref{sec 2: holomodel} can be written as follows 
\begin{subequations}
    \label{BoundaryConditions}
    \begin{align}
    \tilde{A}|_{u=0} &=1 \,\,,  \label{BoundaryCondition_a}\\[1mm]
\tilde{\Sigma}|_{u=0} &=1 \,\,, \label{BoundaryCondition_b} \\[1mm] 
\phi |_{u=0} &=0 \,\,,  \label{BoundaryCondition_c}\\[1mm]
\nu_\phi|_{u=0} &= 1 \,\,, \label{BoundaryCondition_d}\\[1mm]
\tilde{A}|_{u=1} &=0 \,\,,  \label{BoundaryCondition_e}\\[1mm]
\nu_A |_{u=1} &= - 4 \pi T \,\,, \label{BoundaryCondition_f}\\[1mm]
\tilde\Sigma_{u=1} &= \left( S/\pi \right)^{1/3}\,\,.
\label{BoundaryCondition_g}
    \end{align}
\end{subequations}
These conditions are imposed by a combination of the behaviour the metric functions must follow at the boundary (AdS asymptotics at $u=0$, see \eqref{UVexpansion}) and at the horizon at $u=1$, properties of the boundary theory such as its scale and the conformal dimension of its operator deforming the UV fixed point, and the relationship between the metric functions at the horizon and the thermal states at the boundary given in \eqref{temperature_entropy}.

In summary, in the direct problem we choose a value of the scalar field at the horizon $\phi_H$, we obtain the numerical solution of the black brane, and we read off the entropy and the temperature. In contrast, in the inverse problem we start from the equation of state $S(T)$ and use it as a set of BCs to try to recover a potential $V(\phi)$ that fits them all simultaneously.

\subsection{PINNs implementation} \label{subsubsec: NN setup and procedure}
In the following subsections we will present the PINNs-based strategy that was used to improve on the results from our previous work and to probe the false vacuum regime. 
We will take the same base setup that worked for boundary data corresponding to crossovers and to $2^{nd}$-order and $1^{st}$-order phase transitions, and extend it with a number of novel techniques and modifications. These improvements ultimately allow the NNs to overcome the significant difficulties introduced in previous sections associated with inverting potentials in the novel physical regimes we wish to explore.
For extensive details on general aspects of the original setup, such as sampling of the points on the $S(T)$ or the independent variable generator, we refer the reader to \cite{Bea:2024xgv}. Below, we will simply present the improvements and novel techniques used on top of the aforementioned base setup, highlighting the reasons why they are necessary to probe the false vacuum regime.

\subsubsection{Architecture and activations}

\begin{figure}[h!]
    \centering
    \includegraphics[width = \textwidth]{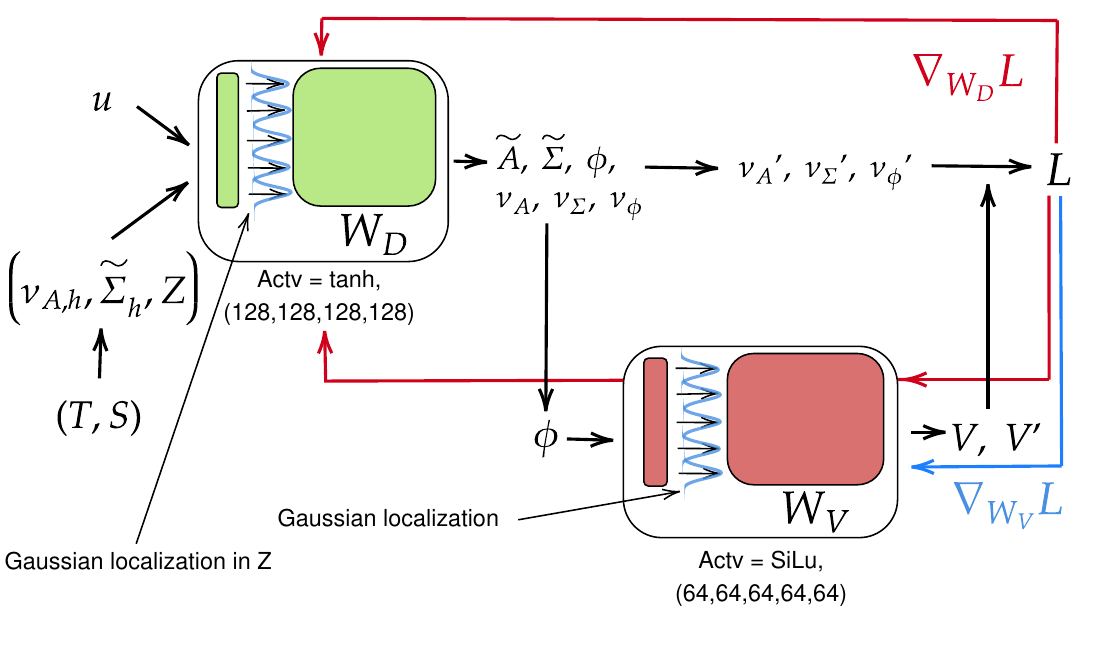}
    \caption{\small Neural Network setup. Novel features include the affine input parameter $Z$, Gaussian localization in $Z$ in the NN-Solver nets and a setup with larger, deeper nets.}
    \label{fig:NN_arch}
\end{figure}

The architecture used for this problem is shown in Fig.~\ref{fig:NN_arch}. The pipeline is comprised of two different  NNs. The first one, NN-Solver, is composed of 6 different nets, each one outputting one of the 6 bulk metric function solutions to the DEs in \eqref{1s_order_ODEs} as outputs, that is, the functions $\tilde{A}, \tilde{\Sigma}, \phi$ and their corresponding first-order derivatives $\nu_A, \nu_{\Sigma}, \nu_{\phi}$. As inputs to these nets, along with the independent variable $u$ and the boundary data $(T,S)$, we will also give a novel input parameter $Z$ to the nets that characterizes the position of each particular sampled point along the $S(T)$ that is used as an input. The numerical computation carried out to obtain this parameter is explained below in subsection \ref{sec: Affine parameter}. In order to give more freedom to the nets in these stiffer regimes, these will have 4 hidden layers with 128 neurons each.

The second network, NN-V, will be solving for the scalar potential $V(\phi)$, taking as input the solution $\phi(u)$ coming from the NN-Solver nets. Once again, now this net has been given more parameters in the form of 5 hidden layers with 64 neurons each. This is particularly important in the cases attempted in this paper, as the aforementioned large separation of scales between the extrema of the potentials to be recovered (see Fig.~\ref{fig:Potentials}) requires an enhanced freedom in the nets if one hopes to accurately capture the defining physical features. However, it is equally important to note that over-increasing the number of available parameters in NNs can lead to overfitting, where the model additionally learns noise and outliers in the data such that it performs poorly on unseen data later on. This particular architecture presented here was determined through trial and error within the explored regime of the inverse problem at hand, adapting proven configurations from the inverse problems attempted in our previous study. In summary, we have:

\begin{align*}
    &[\text{NN-Solver}]_{i} : [128,128,128,128]_{i}, \text{ with } i = 1, \ldots, 6; \\&
    \hspace{11mm}\text{NN-V} : [64,64,64,64,64].
\end{align*}

As previously done, the activation functions used will be the tanh$(x)$ for NN-Solver nets and the \texttt{SiLu} function for the NN-V net, where \texttt{SiLu}$(x) = x(1+e^{-x})^{-1}$.

\subsubsection{Affine parameter}
\label{sec: Affine parameter}
A novel key feature of our implementation is the addition of the affine parameter of the boundary $S(T)$ curve. There are two main advantages that come with introducing this parameter; firstly, it allows us to break the degeneracy between the false and the true vacuum points in the curve and, secondly, it makes it possible to sample the curve uniformly. This in turn allows the PINNs setup to better learn how the obtained solutions change as one moves along the $S(T)$ input parameters.

As was previously explained in Section \ref{subsec: setup}, the QFTs whose dual potential we wish to recover  are characterized by the presence of a  state that has a finite energy density for a vanishing temperature, a so-called false vacuum, as opposed to the true vacuum state of the theory in which the energy density and the temperature vanish simultaneously (see Fig.~\ref{fig:EofT}). However, when one translates energy density to entropy density, both the false vacuum and the true vacuum of the theory are mapped to the point $S = 0$ and $T = 0$. Despite the fact that these two points are very close to each other in phase space, the solutions to the ODE system are very different. This makes the feature problematic for the PINNs setup, since the unknown functions are NNs that depend parametrically both on the entropy and the temperature and hence will be unable to break the degeneracy between the false and the true vacuum. By introducing the affine parameter of the $S(T)$ curve as an additional input to the NN model, the false vacuum will possess a smaller value for the affine parameter than the true vacuum, therefore breaking the degeneracy between these two points when feeding them as inputs to the model.

Furthermore, as mentioned earlier the second advantage given by the affine parameter is that it allows us to uniformly sample the points along $S(T)$. Once this parameter is computed, one can choose a set of evenly-spaced points along the affine parameter direction. In this way, we ensure that each region of the curve has the same weight in terms of number of points contained when training the model.

The numerical computation of the affine parameter is quite standard. We start with a list of points $\{T_i,S_i\}$. In order to compute it, we first need to choose a particular parametrization of the curve $\{T(i),S(i)\}$. We use the positional index as an auxiliary parameter of the curve. With this, we can interpolate both the temperature and the entropy as a function of this parameter. Once this is done, the affine parameter can be computed as the norm of the tangent vector to the curve
\begin{equation}\label{eq: affine parameter}
    Z(s) \,=\, \int_0^{s}\sqrt{\left(\frac{dT}{di}\right)^2 + \left(\frac{dS}{di}\right)^2}\, di,
\end{equation}
where the integration constant is chosen such that the high temperature points correspond to small values of the affine parameter. Finally, this parameter is normalised to $Z \in [0,1]$ and is uniformly sampled in this interval. This sampling will generate an evenly distributed sampling of points $\{T(Z),S(Z)\}$ which will then be given to the neural network as inputs, together with the independent variable $u$ and the affine parameter $Z$.

\subsubsection{Gaussian localization}
\label{subsubsec: gaussian localization}

A key technique to approach the inverse problem at hand, \textit{Gaussian localization} (GL), was introduced in our previous work \cite{Bea:2024xgv}. We will now first review this aspect, providing novel insight on the intuition behind it, and will then detail how it was further implemented on the aforementioned affine parameter, as displayed in Fig.~\ref{fig:NN_arch}.

The GL technique gives the NN the ability to capture a sense of \textit{locality} in some input parameter to the net (in our case, the independent variable or the affine parameter). Specifically, it allows for a tuning of the parameters of the NN that introduces modifications of the solution only in certain regions around a specific value of the input parameter while leaving solutions outside of this region almost unchanged.

The GL technique is implemented as follows. First, one selects the input parameter $x$ that will undergo GL. Then, a gaussian window centered at $\mu_i$ with variance $\sigma_i$ is applied after the first linear transformations of the input parameter (one per neuron of the layer), $w_i\,x + b_i$, before the non-linear activation is applied:

\begin{align}
        x \rightarrow \boxed{\text{Linear Neuron}_i}& \rightarrow  (w_i \, x + b_i)\cdot \text{Exp}\left[-\frac{(w_i\, x + b_i - \mu_i)^2}{4\sigma_i^2}\right] \notag \\
        & = (w_i \,x + b_i)\cdot \text{Exp}\left[-\frac{\left(x+\frac{b_i}{w_i} - \frac{\mu_i}{w_i}\right)^2}{4\left(\frac{\sigma_i}{w_i}\right)^2}\right] \notag \\
        & = (w_i\, x + b_i)\cdot \text{Exp}\left[{-\frac{\left(x - \mu_i'\right)^2}{4\left(\frac{\sigma_i}{w_i}\right)^2}}\right]         \label{eq: gaussian localization} \\
        \text{where} \ \ \ i=1,\dots,\#\text{ neurons in layer 1.} \notag
\end{align}

The first line above is the transformation as shown in \cite{Bea:2024xgv}, while the second and third lines are a mathematical rewriting that makes the GL effect more explicit, where the weight $w_i$ is reabsorbed by dividing the bias $b_i$ and the mean $\mu_i$ together into a new variable $\mu_i'$. The values of the different $\mu_i'$ can be chosen to be either fixed or learnable parameters. In the third line, one can see why this construction works. First, note that the response propagated through the network is only sensitive to the weights and biases that make $w_i\,x + b_i \sim \mu_i'$. Thus, the NN parameters $w_i$ only affect the output solution in a neighbourhood of the input $x\in \mu_i'\pm\sigma_i/2$. 

Furthermore, the fact that the factor $\sigma_i$ is normalized by the weight $w_i$ (while keeping $\sigma_i$ fixed, as detailed below) solves a problem that arises if one naively applies a window $\text{Exp}\left[-\frac{(\phi-\mu_i)^2}{4\sigma_i^2}\right]$ with a learnable value of $\sigma_i$. In the naive case, a single neuron $i$ can take over the whole first layer (e.g. $\sigma_i\rightarrow\infty$), and thus the localization is lost. However, by normalizing in the way presented above, the case where one neuron would take over the whole layer, namely $\sigma^\text{eff}_i = \sigma_i/w_i\rightarrow \infty$, implies $w_i\ll 1$ which makes the whole response in \eqref{eq: gaussian localization} close to zero. On the other hand, if the weight $w_i$ is large, the region of $x$ controlled by this weight is very narrow. This is a desirable outcome, as changes in $w_i$ modify the NN solution in regions that are very localized around $x\approx \mu_i$. 

The fixed values of $\sigma_i$ are taken to be of the order of $(\Delta x)^{-1} = (x_i - x_{i-1})^{-1}$ in the sampling of the input variable $x$, so that they cover the whole span of $x$.\\

\underline{\textbf{GL on the NN-V ($x=\phi$)}}:

For the NN-V network that reconstructs the inverse function $V(\phi)$, the GL is applied to the single input variable $\phi$, following the approach of \cite{Bea:2024xgv}. The scale parameters $\sigma_i$ are fixed to values of order $\Delta\phi^{-1}$, where $\Delta\phi$ is defined based on a uniform discretization of $\phi$, even though the actual $\phi$ samples are generated by a separate neural network and are not uniformly spaced. The means $\mu_i$ are kept as trainable parameters, allowing the NN-V sufficient flexibility to explore a wide class of configurations in order to recover $V(\phi)$.\\

\underline{\textbf{GL on the NN-Solver ($x=Z$)}}:

Additionally, in this work we also employ GL on the affine parameter variable $Z$, which enters as an input parameter to the NN-Solver network. In this case, both $\mu_i$ and $\sigma_i$ values are fixed to linearly spaced values between $0$ and $1$ (span of $Z$), and to $(\Delta Z)^{-1}$, respectively.\\

The intuition behind imposing GL on $Z$ is the following: in the space of parameters, the NN-Solver seems to have a better handle on the solution space when it is allowed to modify a solution $\psi_i$ that corresponds to some BCs given by a point in the equation of state $(S_i,T_i)$ while not changing the ones corresponding to points in the $S(T)$ curve that are distant in the physics of the problem.\footnote{Note that different values of $(S,T)$ might be very close in the equation of state but may represent very different physics, thereby having very different values of the affine parameter $Z$ as explained in Sec. \ref{sec: Affine parameter}. GL can help the NN be more ``flexible'' in the learning process, improving a solution corresponding to $(S,\,T,\,Z_1)$ while barely modifying one that corresponds to $(S,\,T,\,Z_2)$, with $Z_1$ and $Z_2$ very different from each other; this occurs, for instance, with states near the false and true vacuum.}

\subsubsection{Additional losses}
\label{sec: addlosses}
Along with the standard loss function involving the squared residuals of the different ODEs that are to be simultaneously solved in this problem, here we will also make use of additional losses which enter the main loss function as defined in subsequent sections. As explained later in detail, these will have a notably smaller weight than the main residual ODE loss term, but will nevertheless play a very important role in guiding the NNs towards recovering the right functional form in each run. Moreover, each of these additional loss terms comprises physical information that is entirely contained in the input $S(T)$ curve, such that they can simply be regarded as a valid exploit of the physics present when one is given an equation of state. Below, we introduce the two main additional loss terms that we adopted for the inversion problem at hand.\\

\underline{\textbf{UV/IR asymptotic form}}:

In our current theoretical setup, the family of equations of state $S(T)$ considered arises from a theory obtained by deforming a UV CFT by a relevant operator $\mathcal{O}$, triggering an RG flow to an IR CFT. The entropy density  interpolates between the thermodynamic behaviour characteristic of the UV CFT at high temperatures and that of the IR CFT at low temperatures. For a conformal theory deformed by a relevant operator, it can be shown that the relationship between the entropy and the temperature at high temperatures asymptotes to
\begin{equation}
\label{eq: CFT behaviour}
    S = c_1 T^3 + c_2 T^{2\Delta-5} + \cdots ,
\end{equation}
where $c_1,c_2$ are numerical coefficients and $\Delta$ is the conformal dimension of $\mathcal{O}$. In our previous work \cite{Bea:2024xgv}, we already used this fact to extract information from the boundary data since at the UV the conformal dimension is set, $\Delta_{UV} = 3$, and hence given any $S(T)$ one can make a fit in the high 
-temperature regime to extract the coefficients $c^{UV}_1,c^{UV}_2$. In our conventions, this gives\footnote{We note that the same equation appeared as Eq.~(3.1) in \cite{Bea:2024xgv}, where a typo caused the second term on the right-hand side to appear with a plus sign instead.} 
\begin{equation}
\label{UV CFT}
    S = \pi^4 T^3 - \frac{3\pi^4}{64}T + \cdots .
\end{equation}

By solving the EKG equations \eqref{Einstein-scalar_equations} perturbatively near the boundary, this form of the entropy density can be shown to fix the value of the scalar potential and its first two derivatives at the AdS critical point $\phi = 0$, which corresponds to the UV fixed point of the dual boundary theory:

\begin{equation}
\label{UV addloss}
   C_0 \equiv V(0) + 3 = 0, \quad C_1 \equiv V'(0) = 0, \quad C_2 \equiv V''(0) + 3 = 0,
\end{equation}
defining the variables $\{C_0,C_1,C_2\}$ that were used to implement these values for the potential as an additional loss in the implementation of the NN model.
Having exploited the information at the UV from the given equation of state, we provided the PINNs model with these boundary conditions for the potential in the form of an additional loss. However, what was not attempted at the time was to similarly extract information at the IR from the equation of state; as it turns out, one can extend this strategy and play the same game by first making a fit at low temperatures to \eqref{eq: CFT behaviour} to find $c^{IR}_1$ and the corresponding conformal dimension $\Delta_{IR}$. Having previously found the value for $c_1^{UV}$ shown in \eqref{UV CFT}, one can use the relation
\begin{equation}
    \left(\frac{L_{IR}}{L_{UV}}\right)^3 = \frac{c_1^{IR}}{c_1^{UV}}
\label{eq: L with c}
\end{equation}
to find $L_{IR}$ given $L_{UV} = 1$.\footnote{Eq.~\eqref{eq: L with c} comes from the thermodynamic relation $\frac{s}{T^3} =\frac{2 \pi^4 L^3}{\kappa^2_5}$, where this relation is true when the theory is conformal at the UV/IR regimes. By equating said regimes, one can find the relation between the different AdS radii and the leading-order coefficient of the conformal theory deformed by the relevant operator.} Then, we can use this value to find the theoretical value of the potential at the IR thanks to the relation   
\begin{equation}
\label{eq: IR addloss V}
    C_3 \equiv V(\phi_{last}) = -\frac{3}{L^2_{IR}}
\end{equation}
shown in \cite{attems2016thermodynamics, Attems:2016tby}, where the last value of the potential is at $\phi = \phi_{last}$. Since the IR is a fixed point, we know that the potential will also have a critical point at $\phi = \phi_{last}$, and so 

\begin{equation}
    \label{eq: IR addloss DV}
    C_4 \equiv V'(\phi_{last}) = 0.
\end{equation}
Lastly, the second derivative of the potential encodes the mass of the bulk scalar field at the IR fixed point via $V''(\phi_{last}) = m^2_{IR}$. The standard AdS/CFT dictionary \cite{Witten:1998qj, aharony2000large} relates the mass of a scalar in $\text{AdS}_{d+1}$ to the conformal dimension of the dual operator via $m^2L^2 = \Delta(\Delta-d)$. Applied at the IR fixed point in our five-dimensional setup, from this we obtain

\begin{equation}
    \label{eq: IR addloss DDV}    
    C_5 \equiv V''(\phi_{last}) = \frac{\Delta_{IR}(\Delta_{IR}-4)}{L^2_{IR}},
\end{equation}
as used in \cite{Attems:2016tby} for this same family of holographic models.\footnote{Note that this is the IR-side analogue of the definition of $C_2$ in \eqref{UV addloss}, which encodes the same relation at the UV fixed point with $\Delta_{UV} = 3$ and $L_{UV}$ = 1.} Once again, since these boundary conditions on the potential were implicitly contained in the given equation of state, we will make use of them in our NN model. Note that these new additional losses will give information about the value of the potential at $\phi_{last}$, as well as its first and second derivatives there, but they contain no information on what this value $\phi = \phi_{last}$ should be; the NN model must still be able to learn this during training. We will employ these additional losses in both the near-false-vacuum and false vacuum regimes, where in the latter this will provide physical information on the global maximum and minimum points, as well as on the local minimum corresponding to the false vacuum. \\

\underline{\textbf{Separating solutions among the two branches}}:

For runs with equations of state in the false vacuum regime specifically, we will also introduce a new additional loss term. Firstly, as explained in previous sections, these potentials include a region from their local minimum point up until a certain numerical critical point whose values have no dual boundary thermal states, as in this region the dual thermal states belong to a different CFT at the boundary. Hence, the value of the scalar field at the horizon, $\phi(u=1) \equiv \phi_H$, will go from the maximum of the potential at $\phi_H= \phi_0 = 0$ to the local minimum at $\phi_H = \phi_\text{FV}$ and should then ``jump'' to some critical value $\phi_H = \phi_c$ such that boundary thermal states that come after the false vacuum state will have $\phi_H$ values larger or equal to this numerical critical point. To impose this physically motivated statement, we can effectively split the equation of state into two branches in order to separate thermal points before and after the false vacuum. The exact procedure for separating the branches will be explained in detail in the subsection below. For the first branch (before the false vacuum, $\phi\in[\phi_0,\phi_\text{FV}]$), we do not impose any additional constraints on the loss besides the ODE residuals and the values of $V$ and its first and second derivatives at the UV fixed point at $\phi_H=\phi_0$ in the form of \eqref{UV addloss}. The value of $\phi_H$ found by the NN for the solution corresponding to the last point in the first branch of $S(T)$ should, after an appropriate amount of training, approximate the position of the minimum of the potential corresponding to the false vacuum, $\phi_H=\phi_{\text{FV}}^\text{NN}$. However, for the BCs coming from the second branch (after the false vacuum, $\phi_H>\phi_\text{FV}$), we will include an additional loss term that forces the corresponding solutions for $\phi(u)$ to have a value at the horizon $\phi_H$ (i.e. at $u=1$) above the value $\phi_{FV}$ found during the training of the first branch. In particular, this term can be expressed as:

\begin{equation}
    \label{eq: addloss above phi_FV}
    L_\textup{FV} =  \sum_{i \in \,2^\textit{nd}\textup{ branch}} \sigma\left( \phi^\text{NN}_{\textup{FV}} - \phi_{H,i} \right) \ \ ,
\end{equation}
where $\sigma$ is the sigmoid function and $\phi^\text{NN}_{\text{FV}}$ corresponds to the position found by the NN of the local minimum of the potential when trained only on points in the first branch, i.e. $V(\phi^\text{NN}_{\text{FV}})=\min\left(V^{\textup{NN}_1}(\phi)\right)$.

Note that this additional loss term will not be employed in runs outside the false vacuum regime, as there the potentials will not have the local extrema features and thus the procedure of splitting the input BCs into two branches will not be necessary.

\subsection{NN setup and procedure}
Having presented the improvements made on the original PINNs-based setup, we will now explain the procedures that will be used to approach the inverse problem in: first, the near-false-vacuum, and second, the false vacuum regimes, highlighting the differences in the implementation of the setup shown in Fig.~\ref{fig:NN_arch}. 

We will define the near-false-vacuum regime as the set of inverse problems lying in the range $0.5808 < \phi_M \le 0.8$. The lower bound represents the potential that has a degenerate critical point, such that for a $\phi_M$ lower than it we enter the false vacuum regime, while the upper bound is the inverse problem with the most pronounced phase transition that was attempted in our previous work \cite{Bea:2024xgv}. In this regime, the pipeline used will be the same that was employed in said study, but we will now adopt the improved architecture presented in the previous subsection along with its aforementioned novel features. More specifically, given some equation of state $S(T)$:

\begin{enumerate}
\item The NN-Solver network, parametrized by weights $W_D$, takes the following discrete inputs:
\begin{itemize}
    \item a collection of thermodynamic data points $(T_i,S_i)$ satisfying the equation of state
    \item a set of radial grid points $u_n \in [0,1]$ representing the independent variable
    \item a set of affine parameter values $Z_i \in [0,1]$ indexing the thermodynamic data points along the curve
\end{itemize}
For each pair $(T_i,S_i)$ and each $u_n, Z_i$ values the network outputs predictions for the bulk fields
\[
\tilde A(u,Z,(T,S)), \quad \tilde \Sigma(u,Z,(T,S)), \quad \phi(u,Z,(T,S)),
\]
together with the auxiliary quantities
\[
\nu_A(u,Z,(T,S)), \quad \nu_\Sigma(u,Z,(T,S)), \quad \nu_\phi(u,Z,(T,S)).
\]
By construction, these functions exactly satisfy the boundary conditions \eqref{BoundaryConditions}.

Each thermodynamic pair $(T_i,S_i)$ specifies a complete bulk geometry. Derivatives with respect to $u$, namely $\nu_A'$, $\nu_\Sigma'$, and $\nu_\phi'$, are obtained analytically through automatic differentiation.

\item The outputted scalar field values 
\[
\phi_{i,n} = \phi(u_n,Z_i,(T_i,S_i))
\]
are then supplied to a second neural network, NN-V, with parameters $W_V$. This network reconstructs the scalar potential by producing predictions for $V(\phi_{i,n})$. Its derivative with respect to $\phi$ is again computed using automatic differentiation.

\item The outputs of both networks are inserted into the system of ODEs \eqref{1s_order_ODEs}. A global loss function $\mathcal{L}$ is defined and minimized with respect to all trainable parameters $W_D$ and $W_V$ via stochastic gradient descent. The loss function takes the form
\begin{equation}
\begin{split}
    \label{loss function}
\mathcal{L}
=
& \sum_{\alpha}\sum_{n}\sum_{i}  \ 
E_{\alpha}\bigl(u_n,Z_i,(T_i,S_i)\bigr)^2 \\
&+ \lambda_{UV}\left(C_0^2+C_1^2+C_2^2\right)
+ \lambda_{IR}^{(1)}\left(C_3\right)^2+ \lambda_{IR}^{(2)}\left(C_4\right)^2+\lambda_{IR}^{(3)}\left(C_5\right)^2,
\end{split}
\end{equation}

where the first term represents the squared residuals of all equations of motion \eqref{1s_order_ODEs}, summed over the equation index $\alpha$, over the radial grid points $u_n$, and over all thermodynamic configurations $(T_i,S_i)$ with their respective affine parameter $Z_i$. The second term enforces the original additional loss controlling the three boundary conditions on the potential at the UV, \eqref{UV addloss}. Finally, the three last terms represent the new additional losses related to the boundary conditions on the potential at the IR, with the terms ${C_3,C_4,C_5}$ defined above in Eqs. (\ref{eq: IR addloss V}--\ref{eq: IR addloss DDV}). In practice, these three conditions were separated into three additional losses as it was found through experimentation that it was beneficial for the model to have the freedom to learn each condition individually.

The hyperparameters $\lambda_{UV}, \lambda^{(1)}_{IR}, \lambda^{(2)}_{IR}, \lambda^{(3)}_{IR}$ control the strength of the additional losses. Rather than activating them abruptly, each coefficient is staged smoothly during training by modulating its base weight value $\bar{\lambda}$ with a sigmoid ramp in the epoch number $n$,

\begin{equation}
\lambda(n) = \bar{\lambda} \, \sigma \Bigl( a (n - n_0) \Bigr)
\label{eq:staging}
\end{equation}

where $\sigma$ is the sigmoid function, $a$ sets the slope of the ramp, and $n_0$ is 
the turn-on epoch at which the coefficient reaches half of its base value. This 
gradual scheduling allows the model to first minimize the ODE residuals unimpededly, with the additional constraints on the potential then being enforced progressively as training 
advances. Through experimentation, this was found to avoid the optimization instabilities caused by a sudden change in the loss landscape, as well as preventing the model from minimizing the additional loss terms at the expense of the ODE residuals early in training, when the bulk solutions are still inaccurate. In this way, the known physical constraints on the potential act as a late-stage refinement of already-consistent solutions rather than as a competing objective from the outset. In our runs we take $a = 10^{-4}$, $n_0 = 10^5$ and base values 
$\bar{\lambda}_{UV} = 1$, $\bar{\lambda}^{(1)}_{IR} = 10^{-2}$ and 
$\bar{\lambda}^{(2)}_{IR} = \bar{\lambda}^{(3)}_{IR} = 10^{-6}$; these choices were found through experimentation to be generally optimal for the training of the models.

The optimization is performed using the Adam algorithm, which augments stochastic gradient descent with estimates of higher-order moments.

\item
The above procedure is repeated over a fixed number of training cycles (epochs). At each epoch, the loss is evaluated to monitor convergence, and training proceeds until the residuals of all equations fall below a prescribed tolerance. Note that the order of the loss at which training will stop will be different for the two regimes explored in this paper; specific details are given in Section \ref{sec 4: results}. 

\end{enumerate}

The training of an NN model inherently involves randomness, stemming both from the use of stochastic gradient descent as well as the random initialization of the learnable weights and bias parameters inside each neuron. Therefore, as done in our previous implementations of this PINNs-based methodology, when running the above pipeline we perform various different runs of order $\sim 10^6$
epochs and then continue the training for the best run of the batch.

For the false vacuum regime, we will also use the pipeline described above but will employ it following a different strategy that will allow us to overcome the challenges associated with recovering potentials in this regime. More specifically, we will adopt a two-step approach that will involve splitting the equation of state into two separate sections and feeding them sequentially to the NN model. The first branch will span points going from high temperatures at the UV to the false vacuum point, while the second branch will go from the first point after the false vacuum to the true vacuum of the curve, forming an elliptical-shaped section as illustrated in Fig.~\ref{subfig:SofT}. The criterion used to split the branches is that we associate points with high $T$ and high $S$, together with points where the energy density is non-zero while the temperature goes to zero (see Fig.~\ref{subfig:EofT}), to the first branch. These points can also be identified from the $S/T^3$ plot (Fig.~\ref{subfig:soverT3}), where the first branch points go to a constant as $T\rightarrow0$ that is larger than the constant $c_1^{IR}$ from the IR fixed point. The rest of the points are associated to the second branch. Having split the points in the equation of state into these two branches, we will firstly proceed to only give input data from the first branch to our PINNs setup. The pipeline followed will be the same as for the near-false-vacuum case presented above, with the notable difference that the loss function \eqref{loss function} will not include the additional loss enforcing the coefficients $\{C_3,C_4,C_5\}$ at the true vacuum, as this first branch does not contain points pertaining to said regime. In this initial training phase, the solutions and the potential are quickly obtained by the NN model to a good accuracy level, since the points in this section of the equation of state represent states going from the UV to the IR akin to a scenario where one has a crossover transition (typically corresponding to values such as $\phi_M = 5$), which is the transition that is most easily dealt with. From this run, a potential spanning $[\phi_0,\phi_{FV}]$ going from an initial maximum point to a local minimum is recovered, the latter corresponding to the recovered false vacuum thermal state at the boundary.

Following this first training step, the weights in the NN corresponding to the first section of the potential will be frozen (expressed as $V_1^\text{frozen}(\phi)$), and this frozen model will now be ``attached'' to a second (new) NN, $V_2^\text{NN}(\phi)$. This second model is then fed the points in the $S(T)$ pertaining to the second branch. As explained in Sec.~\ref{sec: direct problem}, we know from the theory that these solutions pertaining to the second branch should have a value of the scalar field at the horizon of  $\phi_\text{FV} < \phi_c < \phi_H$, and thus the recovered solutions $\phi^\text{NN}$ should span the range $[\phi_c,\phi_{last}]$. Here, $\phi_{last}$ corresponds to the global minimum of the potential (the true vacuum in the equation of state), while $\phi_c$ is the critical value corresponding to the first thermal state in the second branch after the false vacuum. Both values $\phi_c$ and $\phi_{last}$ must be learned by the NN-setup. The region between $\phi_{FV}<\phi_H<\phi_c$ is the section of the potential that corresponds to thermal states that do not pertain to the boundary equation of state and whose potentials exhibit some of the exotic behaviour briefly discussed in Section \ref{sec: intro}, such as the so-called skipping flows. Recovering the presence of this gap from the $\phi(u)$ solutions from each of the branches, along with the values of the potential evaluated within said gap (corresponding to the presence of skipping flows in the potential) is highly non-trivial, as the NN only has information of the potential at the initial and final points of this region, which additionally contains a local maximum. 

\begin{figure}
    \centering
    \includegraphics[width=0.99\linewidth]{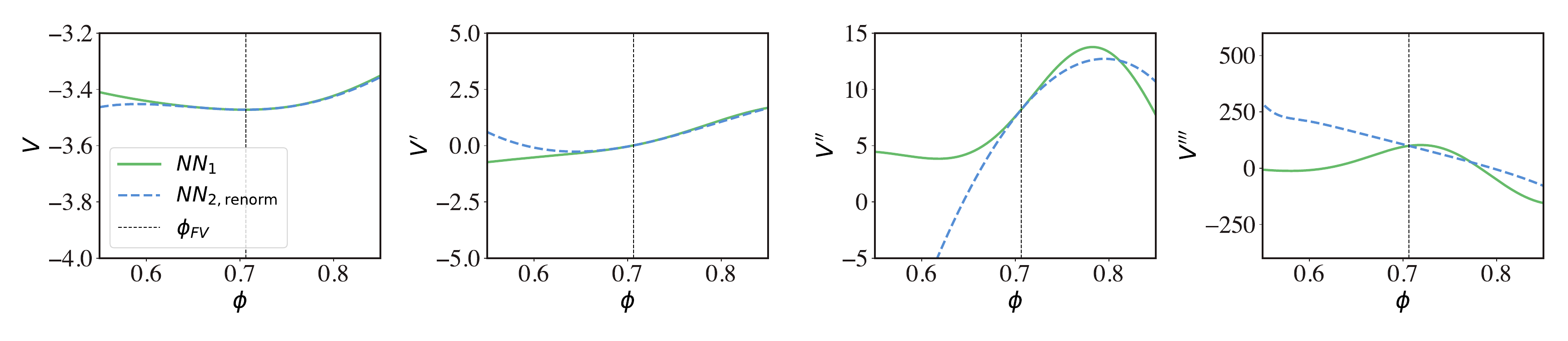}
    \caption{\small Explicit display of the matching-by-construction of value of the two NN-components of the potential $V$ in \eqref{eq: V from matching of both nets} up to the $3^{rd}$ order derivative.}
    \label{fig: matching of V derivatives from both branches}
\end{figure}

The full potential is hence defined to be given by the first NN if $\phi <\phi_{FV}$, and by the second one if $\phi >\phi_{FV}$ through the sigmoid function $\sigma$, in a way that allows the argument above to be realized during training but that does not provide the NN model with additional information besides that which can be obtained from the $S(T)$ curve alone:

\begin{equation}\label{eq: potential combination}
    V(\phi) = V^\textup{frozen}_1(\phi) \cdot \sigma(\phi_\text{FV} - \phi) + V^\text{NN}_{2,\,\textup{renorm}}(\phi) \cdot \sigma(\phi-\phi_\text{FV}),
\end{equation}
where the renormalized potential output to the second, unfrozen NN is:
\begin{align}
\label{eq: V from matching of both nets}
    V_{2,\,\textup{renorm}}^\text{NN}(\phi) & = V_1^\textup{frozen}(\phi_\text{FV}) + \sum_{k=1}^{n-1} \frac{1}{k!} \,(\phi-\phi_\textup{FV})^k \cdot \left.\frac{d^k V_1^\textup{frozen}(\phi)}{d\phi^k}\right|_{\phi=\phi_\textup{FV}}  \notag \\
     & + \frac{1}{n!}\,(\phi-\phi_\textup{FV})^n \cdot V_2^\textup{NN}(\phi) 
\end{align}
This renormalization is such that the value of $V_{2,\,\textup{renorm}}^\text{NN}(\phi  = \phi_\text{FV})$ and the $n-1$ first derivatives evaluated at $\phi_\textup{FV}$ match with those of $V_1^\textup{frozen}(\phi)$. In our pipeline, we set $n=4$. We consider the matching up to $3^{rd}$-order in derivatives of the potential to be enough for our purposes, since derivatives of the potential appear explicitly only up to second order in the Einstein field equations (EFEs) \eqref{1s_order_ODEs}. For an explicit display of the matching-by-construction values of $V$ and its derivatives at $\phi=\phi_\text{FV}$, see Fig.~\ref{fig: matching of V derivatives from both branches}.
We further emphasize that $\phi_\text{FV}$ is a quantity that has been obtained by the first NN, from finding the value of the scalar field at the horizon, $\phi_H=\phi(u=1)$, at the point with $(S,T)\approx(0,0)$ in the first branch of the equation of state $S(T)$, and hence it does not imply additional information besides that which is obtained through running the described pipeline.

\begin{figure}
    \centering
    \includegraphics[width=\linewidth]{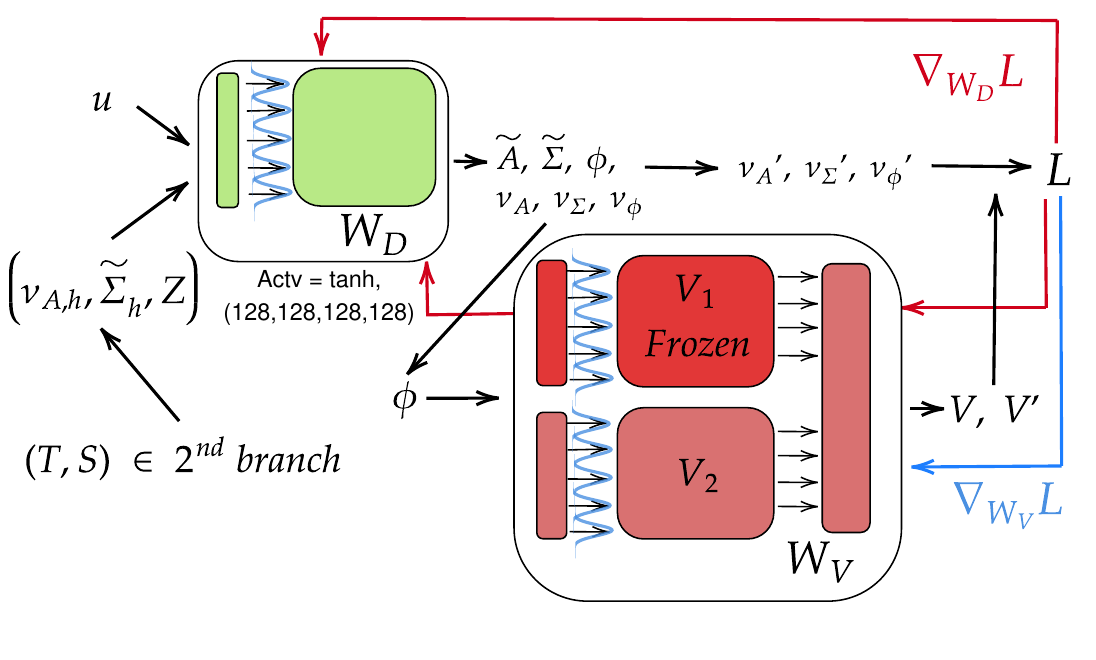}
    \caption{\small NN architecture for the two-branch setup used on false vacuum equations of state. The NN-$V$ network, parametrized by weights $W_V$, is composed of two different models that are trained sequentially; firstly, the weights of NN-$V_1$ are loaded from a pre-training carried out only on input data from the first branch. Then, the NN-$V_2$ and NN-Solver models are trained on the differential equations with only input data from the second branch, while keeping the weights in NN-$V_1$ frozen throughout. The final output of both NN-$V$ models is combined to obtain the final potential $V$ following expression \eqref{eq: potential combination}.}
    \label{fig:NN 2nd branch arch}
\end{figure}

In summary, the pipeline for the false vacuum cases will run with the same general details as the ones presented above for the near-false-vacuum regime, but will proceed in the following steps:

\begin{enumerate}
    \item The NN-Solver and NN-V networks will undergo training using the architecture and activations shown in Fig.~\ref{fig:NN_arch} on only $S(T)$ points belonging to the first branch (UV to false vacuum). The loss function used will be:

    \begin{equation}
    \label{first branch loss function}
    \begin{split}
        \mathcal{L}_1
        =
        \sum_{\alpha}\sum_{n}\sum_{i \in \,1^\textit{st}\textup{ branch}}
        & \ E_{\alpha}\bigl(u_n,Z_i,(T_i,S_i)\bigr)^2 \\
        & +
        \lambda_{UV}\left(C_0^2+C_1^2+C_2^2\right) \\
        & + \lambda_{FV}^{(1)}\left(C^{FV}_3\right)^2+ \lambda_{FV}^{(2)}\left(C^{FV}_4\right)^2+\lambda_{FV}^{(3)}\left(C^{FV}_5\right)^2 
    \end{split}
    \end{equation}
    where the second and third lines are the constraints given by the UV and IR asymptotics of the inputted first branch of $S(T)$, respectively. The values of $C^{FV}_3, \, C^{FV}_4$ and $C^{FV}_5$ are defined as in Eqs. (\ref{eq: IR addloss V}--\ref{eq: IR addloss DDV}), evaluated at $\phi_\text{FV}$ instead of at $\phi_{last}$. Note that the value of $\phi_\textup{FV}$ is learned by the NN-model at this stage of the training. The values of the coefficients in this case are: for the UV, $\lambda_{UV}=10^{-3}$, and for the IR asymptotics of the first branch corresponding to the false vacuum, values begin at $\lambda_{FV}^{(1)}=\lambda_{FV}^{(2)}=\lambda_{FV}^{(3)}= 10^{-6}$, and are increased to $\lambda_{FV}^{(1)} = 10^{-2}$ and $\lambda_{FV}^{(2)}=\lambda_{FV}^{(3)}= 10^{-1}$ after $10^5$ epochs. 

    \item Once trained to recover the potential from the initial maximum point to the local minimum, the weights of the NN-V network are frozen. Then, a new NN-Solver and new NN-V network are initialized. The new NN-Solver will give solutions with a new set of BCs. The new NN-V is attached to the frozen NN-V, and the full model is run only on BCs given by points from the second branch, as shown in the architecture in Fig.~\ref{fig:NN 2nd branch arch}. In this second stage of training, the loss function is modified to:

\begin{equation}
\begin{split}
    \label{second branch loss function}
\mathcal{L}_2
=
\sum_{\alpha}\sum_{n}\sum_{i \in \,2^\textit{nd}\textup{ branch}} & \ 
E_{\alpha}\bigl(u_n,Z_i,(T_i,S_i)\bigr)^2 \\
& + \lambda_{IR}^{(1)}\left(C_3\right)^2+ \lambda_{IR}^{(2)}\left(C_4\right)^2+\lambda_{IR}^{(3)}\left(C_5\right)^2 \\
& + L_{FV} \ \ ,
\end{split}
\end{equation}
including the additional losses at the IR global minimum of the potential $\phi_H=\phi_{last}$ presented in subsection \ref{sec: addlosses} in Eqs. (\ref{eq: IR addloss V}--\ref{eq: IR addloss DDV}), and at $\phi_H=\phi_\text{FV}$ shown in \eqref{eq: addloss above phi_FV}. The value of $\phi_{last}$ is learned by the NN-setup at this stage of training.
    
\end{enumerate}

\section{Results}
\label{sec 4: results}

Here, we present the results obtained by applying the pipeline presented in the previous section to the inversion problem at hand, aiming to recover the bulk scalar potential $V(\phi)$ given an equation of state $S(T)$ in the near-false-vacuum and false-vacuum regimes. 

\subsection{Near-false-vacuum regime ($\phi_M = 0.8$)}

\begin{figure}[t]
     \centering
      \includegraphics[width=0.75\linewidth]{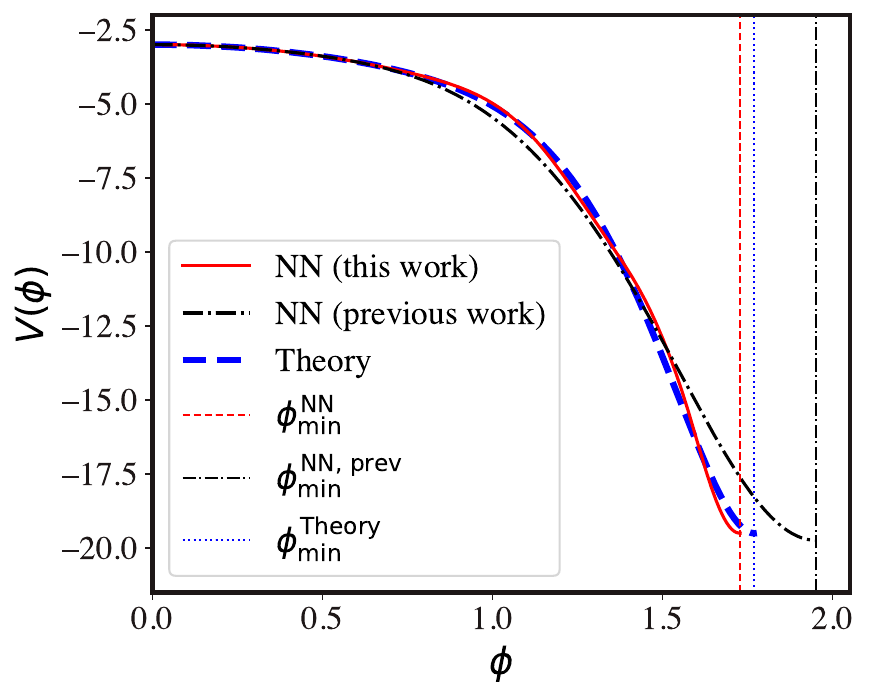}
     \caption{\small Inverted potential in the near-false-vacuum regime for $\phi_M=0.8$. We present results for the NN-pipeline of this work (solid red) compared to the best result obtained in the same case from our previous work \cite{Bea:2024xgv} (dash-dotted black) and the true theoretical function (dashed blue).}
     \label{fig:V(0.8)}
 \end{figure}

To make contact with our previous work \cite{Bea:2024xgv}, we will first present results on the most difficult case attempted there, which being the $\phi_M = 0.8$ case sits within the near-false-vacuum regime. For this problem, we will use the pipeline shown in Fig.~\ref{fig:NN_arch} and described in detail in the previous section. The results for the potential obtained by the NN model are shown in Fig.~\ref{fig:V(0.8)}, where it is displayed against the theoretical potential together with the best result previously achieved in this case in our earlier work. The potential obtained from the NN model in this near-false-vacuum regime, with only minor deviations from the theoretical potential, shows a significant improvement over the one previously obtained with the old NN setup, displaying the strengths of the novel features adopted in our current pipeline. Moreover, to achieve this result the NN model only required training for 2 million epochs, which is less than the 3.5 million epochs of training that the old models underwent to obtain a worse potential. The model was trained until the loss fell to an order of $10^{-6}$, from which point further training was seen to lead to little improvement.

In order to quantitatively evaluate the accuracy of the inverted potential, at first one might simply compare said potential from the NN against the theoretical potential by calculating the relative error (RE) between the two. This comparison can be easily made once a potential is obtained from the NN model by computing:

\begin{equation}
\label{eq: relative error V}
    \delta V(\phi) = \left|\frac{V_{\text{theory}}(\phi) - V_{\text{PINN}}(\phi) }{V_{\text{theory}}(\phi) }\right|.
\end{equation}

\begin{figure}[t]
     \centering
     \begin{subfigure}[b]{0.49\textwidth}
        \centering
        \includegraphics[width=\textwidth]{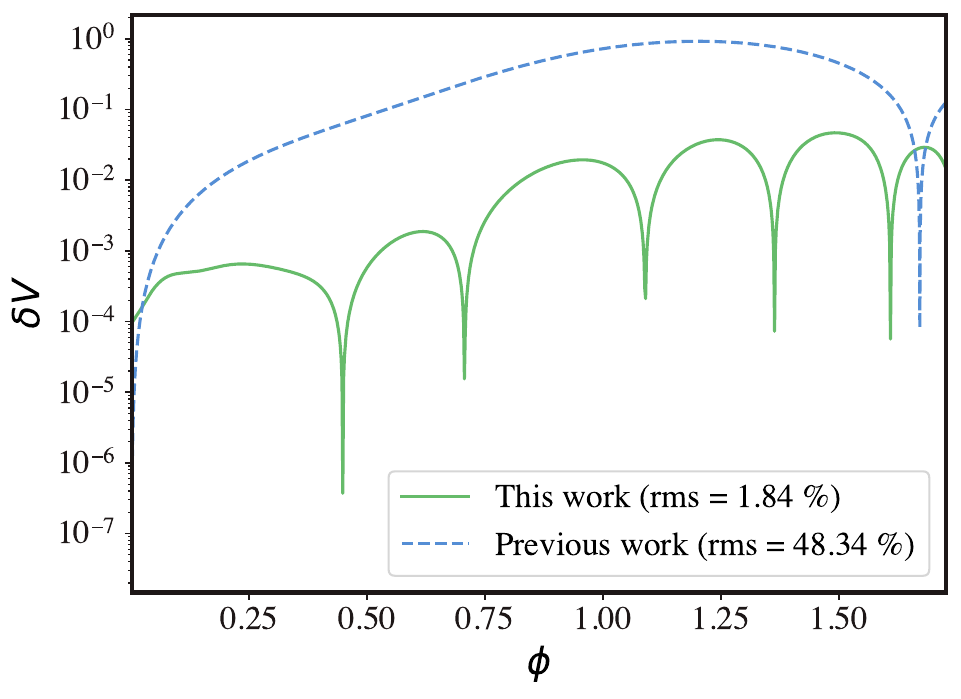}
        \caption{\small }
        \label{fig:RE(0.8) V}
    \end{subfigure}
    \hfill
    \begin{subfigure}[b]{0.49\textwidth}
        \centering
        \includegraphics[width=\textwidth]{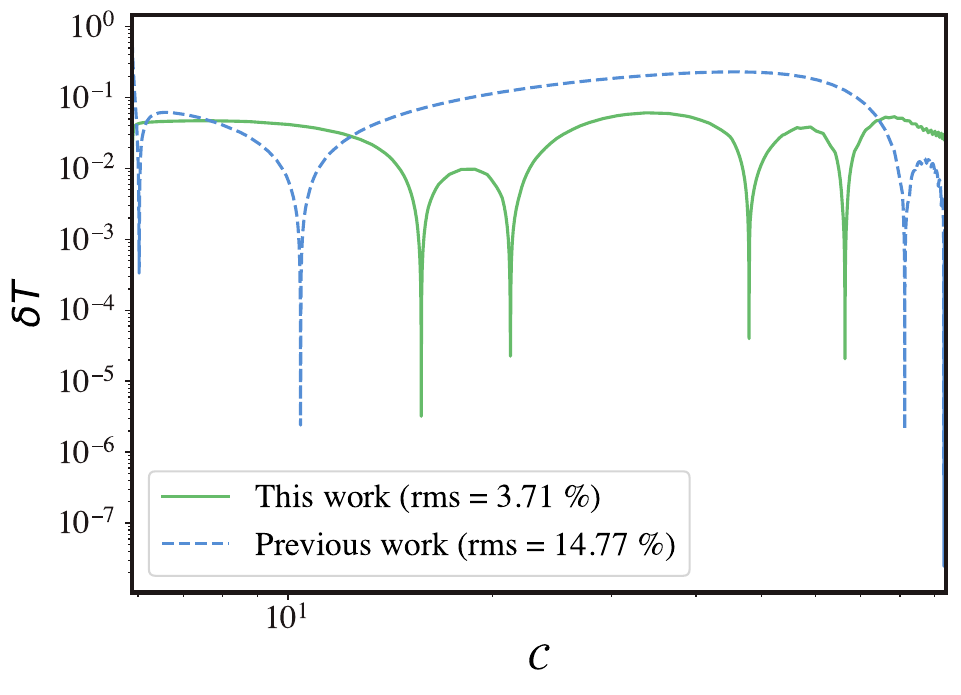}
        \caption{\small }
        \label{fig:RE(0.8) SofT}
    \end{subfigure}

     \caption{\small Reconstruction errors for the near-false-vacuum regime with
        $\phi_M=0.8$. We show results for this work (solid green) and for the
        previous work \cite{Bea:2024xgv} (dashed blue) for the same case, with
        the rms value for each case shown in the legend. (a): RE in the recovered
        potential $V_\text{PINN}(\phi)$ as a function of $\phi$, with an rms
        value of $1.84\%$. (b): RE in the recovered equation of state
        $T_\text{PINN}(\mathcal{C})$, with an rms value of $3.71\%$, where
        $\mathcal{C}\equiv S/T^3$.}
     \label{fig:RE(0.8)}
 \end{figure}

Additionally, we can compute the root mean square\footnote{The root mean square is defined as $\text{rms} (\%)=\sqrt{\frac{1}{n}\sum_{i=1}^n x_i^2}\cdot 100\%$, where $x_i$ is in our case the RE.} (rms) of the RE to obtain a single value that quantifies the error related to the recovered function. These results are shown in Fig.~\ref{fig:RE(0.8) V}, where we can observe the RE to never exceed $4.7\%$ with an rms value of $1.84\%$. Comparing these results with the ones presented in \cite{Bea:2024xgv} for the same case of $\phi_M=0.8$ the old NN-model obtained a maximum value for the RE of around $92\%$ and an rms of $48.34\%$ in its best run.
Hence, we can observe a substantial improvement in terms of error magnitude, as the error in the recovered potential presented in this work shows a factor $\sim20$ improvement in the maximum RE and a factor $\sim26$ in rms with respect to the previous best-case results for $\phi_M=0.8$ in \cite{Bea:2024xgv}. Furthermore, the error quantifications obtained here are generally in agreement with the values obtained in earlier works for easier realizations of higher $\phi_M$ values, which is notable given the increased difficulty of the inversion problem in cases within the near-false-vacuum regime. 

\begin{figure}[t]
    \centering
    \begin{subfigure}[b]{0.49\textwidth}
         \includegraphics[width=\linewidth]{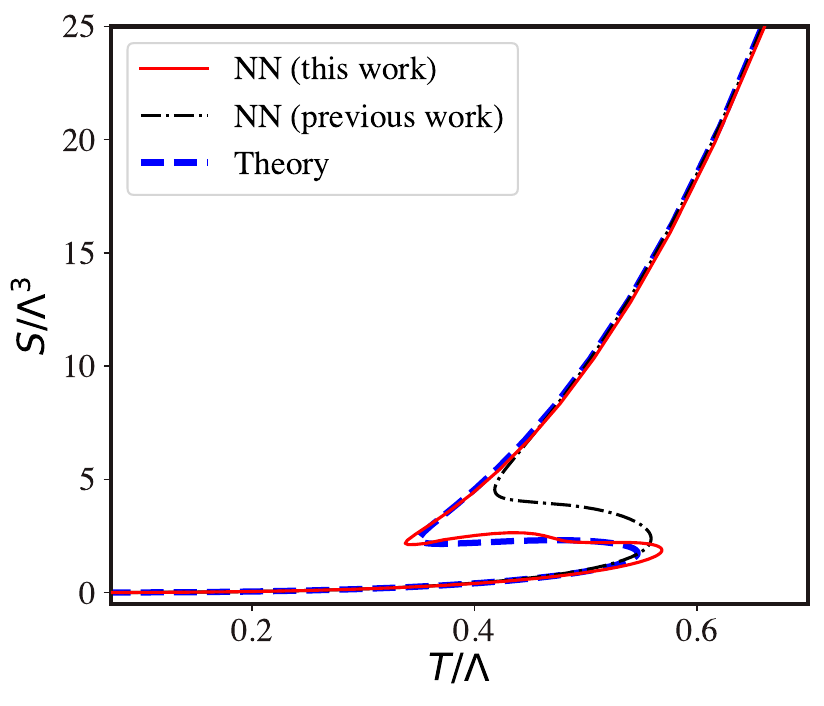}
         \caption{}
     \end{subfigure}
    \begin{subfigure}[b]{0.49\textwidth}
         \includegraphics[width=\linewidth]{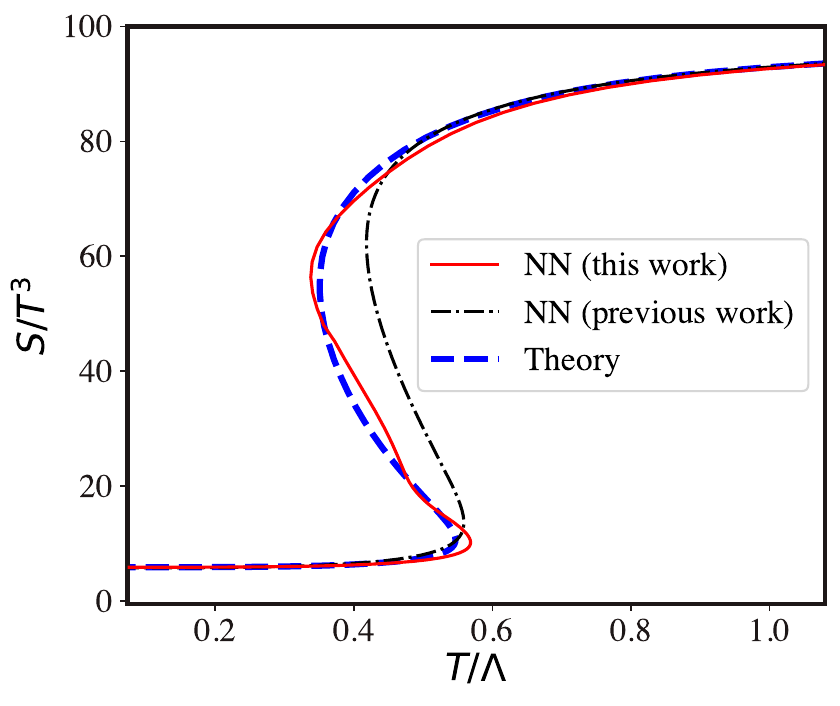}
         \caption{}
     \end{subfigure}
     \caption{\small Equation of state recovered from the inverted potential for $\phi_M=0.8$, presented as $S(T)$ (a) and $S/T^3(T)$ (b). We show the results for this work's NN pipeline (solid red) compared to the input theoretical equation of state (dashed blue), and to the previous work's results \cite{Bea:2024xgv} for the same problem (dash-dotted black).}
     \label{fig:S(T) 0.8}
 \end{figure}

Assessing the quality of the recovered potentials $V_\text{PINN}(\phi)$ as presented above is rigorous as a way of testing our NN-pipeline. However, one could find the error in \eqref{eq: relative error V} unsatisfactory if viewed as a circular argument, since in principle one should be able to use the pipeline presented in this paper for an equation of state obtained in some other manner (for instance, from experimental sources) without requiring a prior theoretical holographic model, $V_\text{theory}(\phi)$. Hence, for any equation of state used in our pipeline, $S_{\text{input}}(T)$, we will take the potential obtained from the NN model, $V_{\text{PINN}}(\phi)$, and use it to solve the direct problem to recover an equation of state $S_{\text{PINN}}(T)$ which we can then compare with $S_{\text{input}}(T)$ to measure the accuracy of the model. Note, however, that this is a very stringent test, since the set of equations being solved both in the direct and inverse problems (namely EFE + KG) is highly non-linear, and thus small errors in the recovered potential $V_\text{PINN}(\phi)$ can be largely amplified when passing through these non-linear equations. The $S_{\text{PINN}}(T)$ curve obtained from carrying out the direct problem using the recovered NN potential is shown in Fig.~\ref{fig:S(T) 0.8}, where the curve $S_{\text{PINN}}/T^3$ is additionally plotted to display the UV and IR behaviours more clearly. Despite the existing deviations from the true equations of state in the middle sections of the curves, which are attributed to the same small deviations that are present in the recovered potential in Fig.~\ref{fig:V(0.8)}, the equations of state are generally in very good agreement with the theoretical input curves.

Following the previous discussion, the RE in the recovered equations of state is calculated using:

\begin{equation}
\label{eq: relative error S(T)}
    \delta T(\mathcal{C}) = \left|\frac{T_{\text{input}}(\mathcal{C}) - T_{\text{PINN}}(\mathcal{C}) }{T_{\text{input}}(\mathcal{C}) }\right|,
\end{equation}
where the function $\mathcal{C}(T)\equiv S/T^3$. The RE in the recovered equation of state is calculated using $T(\mathcal{C}) $ instead of the usual $S(T)$, $T(S)$ or $\mathcal{C}(T)$ due to the former not being multi-valued, unlike the latter three. The feature of multi-valuedness of $T(S)$ is more pronounced in the false-vacuum regime (i.e. for $\phi_M=0.55$). These multi-valuedness arguments can be seen in Figs. \ref{subfig:SofT} and \ref{subfig:soverT3}.

The resulting RE is shown in Fig.~\ref{fig:RE(0.8) SofT}, where it can be seen to never exceed a value of around $6.1\%$ and has an rms value of $3.71\%$. Comparing these values once again to those presented in \cite{Bea:2024xgv} for the same case, there a maximum RE of around $62\%$ and an rms of $14.77\%$ were obtained. Therefore, the results presented in this work represent improvements by approximate factors of 10 and 4 in the maximum RE and rms values, respectively, for the same quality tests performed. Hence, the RE for this recovered phase transition in the near-false-vacuum regime is comparable to that of the phase transitions recovered in \cite{Bea:2024xgv} in simpler regimes. This is a notable result, as the number of additional difficulties related to the large separation of scales that come with recovering potentials in this regime made it inaccessible to the PINNs-based model up to now.

\subsection{False vacuum regime ($\phi_M=0.55$)}

\begin{figure}[t]
\centering
    \includegraphics[width=0.875\linewidth]{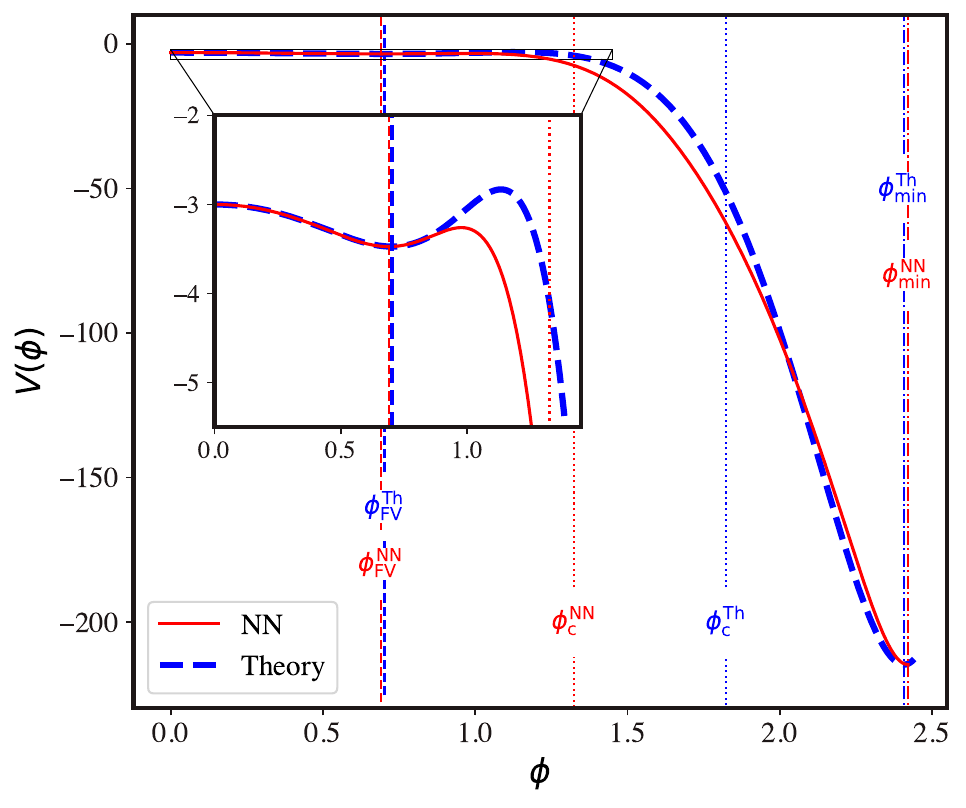}
     \caption{\small Recovered potential (solid red) compared to the theoretical potential (dashed blue) for $\phi_M=0.55$. The potential corresponding to the first branch (FB) ranges from the global maximum to the local minimum denoted as FV (False Vacuum), marked by blue and red dashed lines corresponding to its theoretical and NN-recovered positions, respectively. The strong agreement in the two lines shows the potential for the FB is recovered to a high accuracy. The potential found for the thermal states in the second branch (SB) spans the region from the local to the global minimum, approximately tracing out the correct local maximum in between. The theoretical and NN-recovered  values for $\phi_c$ (dotted) and $\phi_\text{min}$ (dash-dotted) are also shown.}
     \label{fig:0.55 potential}
\end{figure}

To explore the false vacuum regime, we chose to tackle an equation of state well within this domain, namely that corresponding to a dual $\phi_M = 0.55$ theory. Although this choice implies that the dual potential will have a larger separation of scales between its characteristic local and global stationary points, complicating the inversion problem further, it also will have a more pronounced local maximum. Since this is a main feature in the potential which is not encoded by thermal states at the boundary data, it will prove useful for the NN model to clearly distinguish it from the local minimum when recovering the potential. Through experimentation, we observed that if on the other hand one attempts the inversion task on theories closer to the false vacuum threshold, $\phi_M \simeq 0.5808$, the resulting local extrema are very close to being degenerate to a single inflection point, which is a highly fine-tuned feature that is very difficult for the model to accurately find.

 \begin{figure}[t]
     \centering
     \begin{subfigure}[b]{0.49\textwidth}
        \centering
        \includegraphics[width=\textwidth]{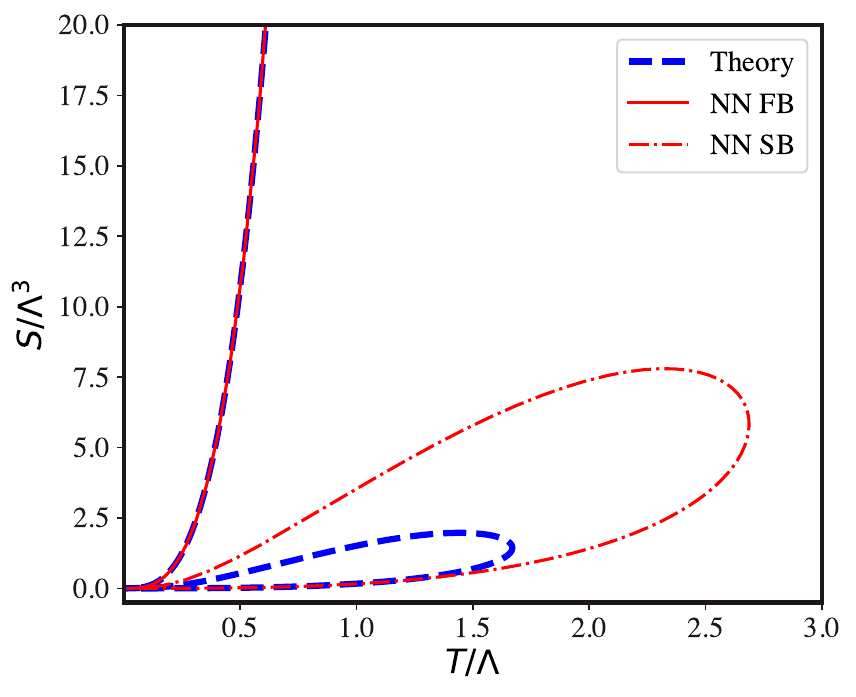}
        \caption{\small }
    \end{subfigure}
    \begin{subfigure}[b]{0.49\textwidth}
         \centering
          \includegraphics[width=\textwidth]{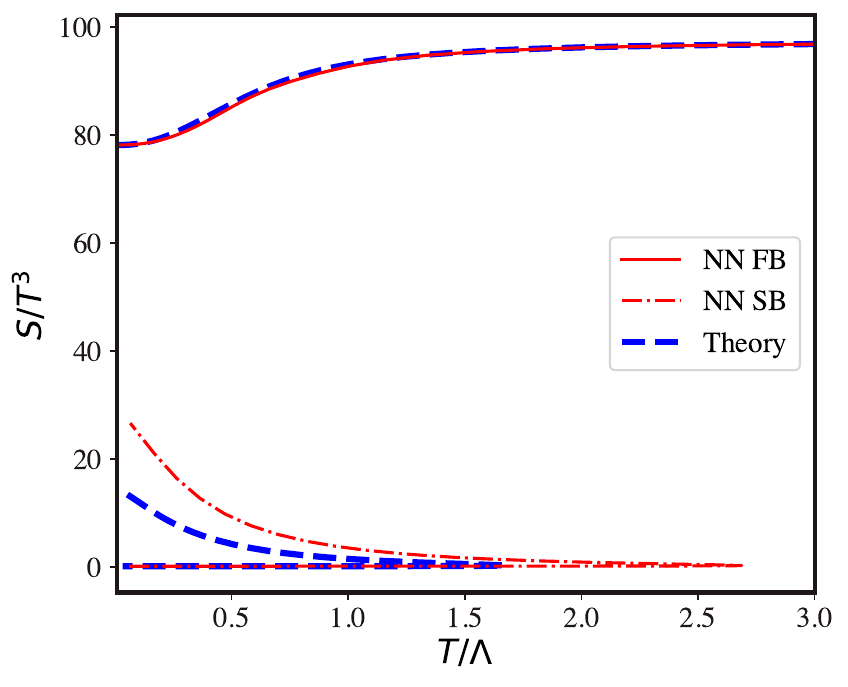}
      \caption{\small }
    \end{subfigure}
    \caption{\small Equation of state recovered from the inverted potential for $\phi_M=0.55$. \textup{(a)}: $S(T)$ for the first branch (FB) (solid red) and second branch (SB) (dash-dotted red), versus the theory (dashed blue). (b): $S/T^3(T)$ for the FB (solid red) and SB (dash-dotted red) versus the theory (dashed blue). The morphology of the recovered curve matches the inputted $S(T)$, and the false and true vacuum points are correctly recovered. }
    \label{fig:0.55 EoS}
 \end{figure}

\begin{figure}[t]
    \centering
    \begin{subfigure}[b]{0.49\textwidth}
        \centering
        \includegraphics[width=\textwidth]{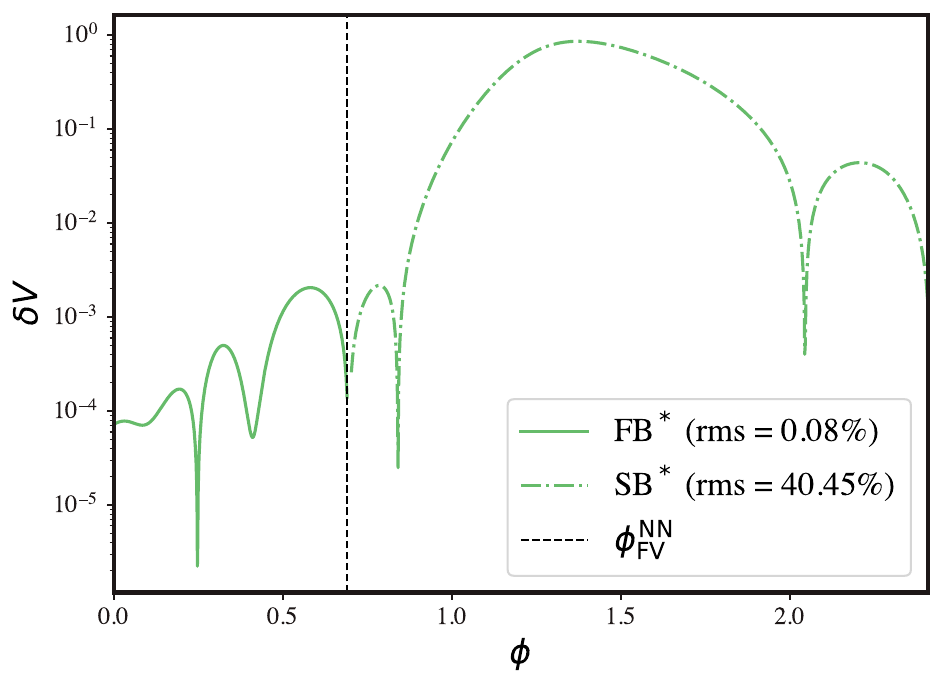}
        \caption{\small }
        \label{subfig: RE for V}
    \end{subfigure}
    \begin{subfigure}[b]{0.49\textwidth}
        \centering
        \includegraphics[width=\textwidth]{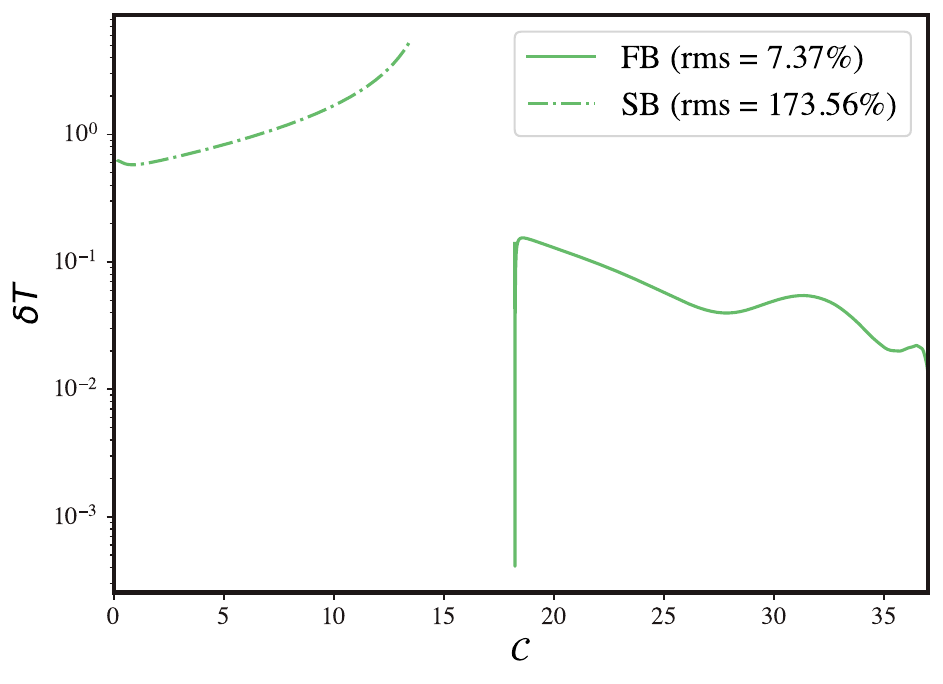}
        \caption{\small }
        \label{subfig: RE for S(T)}
    \end{subfigure}
    \caption{\small Reconstruction errors for the false vacuum regime with
        $\phi_M=0.55$. (a): RE in the recovered potential $V_\text{PINN}(\phi)$ as a
        function of $\phi$, with rms values of $0.08\%$ in the FB and $40.45\%$
        in the SB. The dashed black line indicates the false vacuum point which
        divides the first and second branches represented in solid and dash-dotted
        green lines, respectively. (b): RE in the recovered equation of state
        $T_\text{PINN}(\mathcal{C})$, with rms values of $7.37\%$ in the FB
        (solid green) and $173.56\%$ in the SB (dash-dotted green), being
        $\mathcal{C}\equiv S/T^3$. The FB curve has been shifted to lower values by
        $\Delta\mathcal{C}=60$ for visualization purposes.}
    \label{fig: REs for phiM 0.55}
\end{figure}

For this run, the NN model was trained for 27 million epochs, at which point the loss fell to an order of $10^{-5}$; further training from then was seen to lead to little improvement. The results for the potential found by the NN are shown in Fig.~\ref{fig:0.55 potential}. First, it can be observed that the potential recovered from boundary data points corresponding to the first branch (FB), that is the section of the potential from its global maximum to its local minimum, is recovered to a very high accuracy, as shown by the strong matching between the theoretical and numerical positions of the local minimum (which corresponds to the false vacuum state), as well as the overall precise tracing of the theoretical potential. The RE in this region of the potential corresponding to the FB, namely $[\phi_0, \phi_\text{FV}]$, is below $1\%$, as can be seen in Fig.~\ref{subfig: RE for V}. 

For boundary data from the second branch (SB), the NN model is able to correctly find the local maximum feature as well as the global minimum point. The potential mostly deviates in the region whose dual thermal states are not part of the boundary data, which is to be expected, but remains remarkably close in morphology, content and general location of characteristic features to the true potential even with the considerable separation of scales present in the current theory. When it comes to the equation of state recovered (Fig. \ref{fig:0.55 EoS}) using this using this learned potential, one can similarly see that it matches the original boundary curve to a good relative precision, including its crucial points near the origin corresponding to the false and true vacuum states. The pointwise REs for both the full learned potential and recovered $S(T)$ are shown in Figs.~\ref{subfig: RE for V} and \ref{subfig: RE for S(T)}, respectively. Here, it is important to note that the reported error values appear large precisely in the SB region without boundary input data not because of poor reconstruction, but as an artifact of normalization: both quantities are divided pointwise by a reference value ($|V_\text{th}|$ and $|T_\text{input}|$, respectively) that becomes small near the local extrema and the $(S,T)\approx(0,0)$ vacua, while the absolute deviations there are simultaneously largest owing to the absence of direct thermal data. The combination of small denominator and largest absolute error inflates the error ratio even where the recovered functions track the true ones closely.

\begin{figure}[t]
    \centering
    \begin{subfigure}[b]{0.53\textwidth}
        \includegraphics[width=0.99\linewidth]{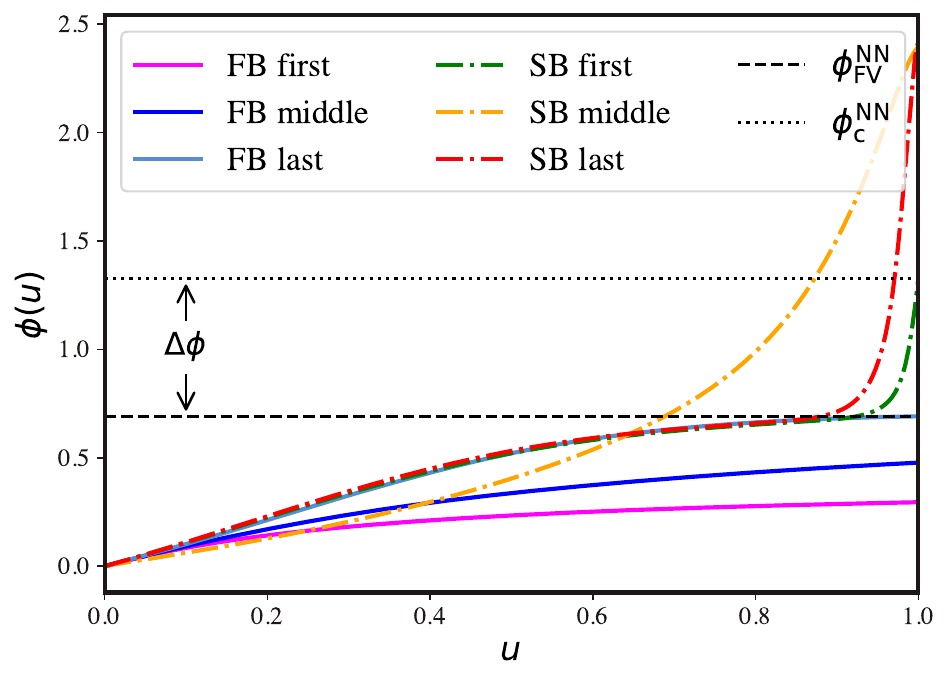}
        \caption{\small }
    \end{subfigure}
    \begin{subfigure}[b]{0.46\textwidth}
        \includegraphics[width=0.99\linewidth]{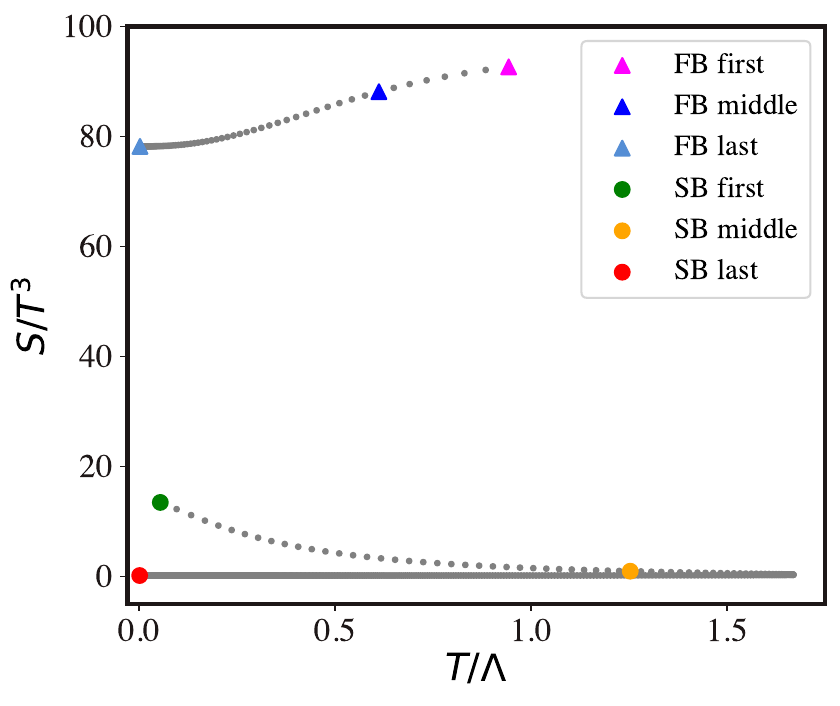}
        \caption{\small }
        \label{subfig: soverT3 special points}
    \end{subfigure}
    \caption{\small (a): Solutions for the scalar field $\phi(u)$ for the case with $\phi_M = 0.55$. The solutions below the dashed horizontal black line correspond to BCs from the FB of the $S(T)$ curve (solid lines), while the ones above it correspond to points in the SB (dash-dotted lines). The NN-model is able to recover the feature of the existence of a gap $\Delta \phi$ in the values $\phi(u=1) = \phi_H$. This is a defining feature in false vacuum cases, where the region of values of the $\phi_H$ where $\phi_\text{FV}<\phi_H<\phi_c$ corresponds to thermal states of a different dual theory. The labels ``first'', ``middle'' and ``last'' indicate the correspondence of the BCs in the equation of state (panel (b)) with the obtained solutions. (b) Equation of state with BC points in the FB and SB corresponding to solutions of panel (a).}
    \label{fig:0.55 phi}
\end{figure}

With regards to the bulk metric field solutions obtained by solving the coupled ODE system, which the NN model must do simultaneously in order to reconstruct the potential, evidence of the false vacuum can also be observed directly from the learned profiles of $\phi(u)$ shown in Fig.~\ref{fig:0.55 phi}. Looking at the figure, one can see the solutions from boundary input data coming from the FB only reach up to a certain maximum value of $\phi(u=1) = \phi_H$, represented by a dashed black line, captured by the solution in light blue whose dual thermal state is that of the false vacuum $\phi_\text{FV}$. These are the same values that were used in the additional loss defined in \eqref{eq: addloss above phi_FV}. Afterwards, there is a significant gap in the solution space quantified by $\Delta\phi \simeq 0.63$
between this solution and the subsequent one for the $\phi(u)$ dual to the first thermal state from the SB, whose value at the horizon is taken to define the quantity $\phi_c$. This parameter is found by the NN-model as a by-product of the training procedure, and for this run it was found to be $\phi_c \sim 1.32$. As explained above, this is an expected physical feature of the theory, since solutions with $\phi_\text{FV}<\phi_H<\phi_c$ correspond to thermal states of a different QFT. Nevertheless, it is noteworthy that the NN-model is able to identify this structure, as the thermal states whose dual $\phi(u)$ solutions attain the delimiting horizon values $\phi_\text{FV}$ and $\phi_c$ are parametrically very close in the equation of state and possess very similar $(S,T)\approx(0,0)$ values, despite corresponding to notably different physics in the bulk theory. It is remarkable that these acutely fine-tuned solutions and relevant physical features are recovered by the model when simultaneously solving for the EKG equations for different boundary conditions. We note that this is in part thanks to the introduction of the novel features presented in Section \ref{subsubsec: NN setup and procedure} such as the affine parameter, which helps break the degeneracy between the last point of the FB and first point of the SB (light blue triangle and green circle in Fig. \ref{subfig: soverT3 special points}, respectively).

\section{Discussion}
\label{sec 5: discussion}
In this work, we have presented a natural extension of the PINNs-based model of \cite{Bea:2024xgv} applied successfully to new regimes previously inaccessible to NN methodologies. The main achievements of this study are twofold. First, in the near-false-vacuum regime ($\phi_M = 0.8$), the combination of the novel techniques introduced here, including the affine parameter for uniform curve sampling, Gaussian localization on the affine parameter direction, and the new additional losses enforcing boundary conditions at the IR fixed point yields a substantially improved reconstruction of the scalar potential compared to the results of \cite{Bea:2024xgv,tarancon2025efficient}, where this case represented the boundary of applicability of the PINNs-based methodology. The recovered equation of state, obtained by solving the direct problem with the NN-reconstructed potential, agrees well with the input data.

Secondly, in the false vacuum regime ($\phi_M = 0.55$) we have successfully recovered the bulk potential for the first time from boundary thermodynamic data. This constitutes a qualitative advance over previous results, as the potential in this regime contains genuinely new features that are  only indirectly encoded in the input equation of state. Specifically, solutions with $\phi_H$ in the region of the potential between the local minimum (the false vacuum) and a critical value $\phi_c$ do not correspond to thermal states of the original QFT whose thermodynamic curve $S(T)$ is given as an input to the NN. The fact that the model is nevertheless able to approximately reconstruct this ``hidden'' portion of the potential, including the correct identification of the local maximum, illustrates the constraining power of the Einstein equations when combined with boundary data from both sides of the gap.

A related result concerns the solutions for the bulk scalar field $\phi(u)$. As shown in the previous section, the NN model is autonomously able to find the gap in the scalar field values at the horizon, $\phi_H$, that separates solutions from the first and second branches. The thermal states whose dual solutions have $\phi_H$ values at the boundaries of this gap ($\phi_{FV}$ and $\phi_c$) are parametrically close in the $(S, T)$ plane, yet they correspond to very different bulk solutions. The ability of the model to resolve this near-degeneracy illustrates the usefulness of the affine parameter $Z$, which makes it possible for the NN to clearly distinguish these distinct boundary conditions and obtain this important consistency check.

Another notable feature used is the implementation of the two-branch training strategy employed for the false vacuum case. By first training on boundary data from the UV to the false vacuum, the model quickly and accurately recovers the section of the potential from the global maximum to the local minimum, a problem that is structurally similar to the crossover case in \cite{Bea:2024xgv}. Freezing the weights of this model and then training a new NN-Solver on the second branch allows the model to focus entirely on the more challenging portion of the potential, from $\phi_c$ to the global minimum. The smooth stitching of the two NN-V models via a sigmoid transition and Taylor matching at the false vacuum point ensures continuity of the reconstructed potential and its derivatives. The generality of this decomposition strategy may prove useful more broadly for inverse problems in which the solution space has natural separation of scales or hierarchies.

Several aspects of the current results invite further improvement. For the $\phi_M = 0.55$ case, while the potential reconstruction in the ``hidden" region between $\phi_{FV}$ and $\phi_c$ captures the correct qualitative features, it equally shows quantitative deviations from the theoretical potential. This is to be expected: the NN has no direct thermodynamic data in this region and must rely solely on the differential equations and the information available at the endpoints. It would be valuable to explore whether additional physical constraints could be incorporated as additional losses to help further constrain the potential.

Another open direction concerns theories whose $\phi_M$ is close to (or is) the critical threshold $\phi_M \simeq 0.5808$, where the two local extrema are very nearly (or become) a degenerate critical point at which both $V'$ and $V''$ vanish. This regime is physically interesting because such theories provide models of out-of-equilibrium, self-sustained inflation \cite{Casalderrey-Solana:2025cdy}. Through experimentation, we have observed that this is also a particularly challenging limit, as the fine-tuned structure in the small-scale features is difficult for the model to resolve. This is especially true when the potential exhibits a degenerate critical point, since in that case the NN model must identify a feature consisting of a single point rather than an extended region of the potential. A systematic study of the model's performance across a finer grid of $\phi_M$ values would help delineate the current boundaries of the method and identify where further technical advances are needed.

Looking ahead, the results of this work open the door to several compelling applications, such as the study of false vacuum decay via bubble nucleation and the subsequent bubble dynamics in the holographic framework. Once a gravitational theory has been reconstructed for a theory of interest whose thermodynamics is known, the full machinery of holography becomes available for the study of far-from-equilibrium dynamics, including transport coefficients, thermalisation, and the approach to equilibrium.

Finally, from a methodological perspective it is worth highlighting that the potential capabilities of the techniques developed and implemented in this work are not restricted to a holographic context. Instead, they address generic difficulties that arise when using PINNs to solve inverse problems involving large physical hierarchies, degenerate boundary conditions, and solution spaces with gaps or discontinuities. We expect them to be applicable to a broader class of inverse problems involving highly nonlinear differential equations.

\acknowledgments

We acknowledge financial support from the ``Center of Excellence Maria de Maeztu 2025--2029'' award to the ICCUB, grant CEX2024-001451-M, funded by AEI/10.13039/501100011033. This work was also partially financed by a grant from the Simons Foundation (00017375, RJ). Partial funding for the work of RJ, PS, PTa and PTe was provided by project PID2022-141125NB-I00. The work of PS was supported by a grant from the ``la Caixa'' Foundation (ID 100010434), under the fellowship code LCF/BQ/DI23/11990074.
PTa is supported by the project ``Dark Energy and the Origin of the Universe'' (PRE2022-102220), funded by MCIN/AEI/10.13039/501100011033. PTe is supported by the FPI predoctoral grant PREP2022-000507, funded by MCIN/AEI/10.13039/501100011033 and by ESF+. DM acknowledges financial support from Grant No. PID2022-136224NB-C22 from the Spanish Ministry of Science, Innovation and Universities, and from Grant No.~2021-SGR-872 funded by the Catalan Government.
This research is also funded by the European Union (ERC, HoloGW, Grant Agreement 
No.~101141909). Views and opinions expressed are, however, those of the authors only and do not necessarily reflect those of the European Union or the European Research Council. Neither the European Union nor the granting authority can be held responsible for them.

\section*{AI and LLM usage disclaimer}
Standard AI programming tools were used during the development of the code that yielded the results shown in this paper. These were used to increase efficiency in tasks such as debugging and modifying existing code. Large language models were used to assist in editing and polishing the manuscript, but they did not contribute to the scientific content or results.

\section*{Code Availability Statement} 
This article has associated code in a code repository.
Available at: \href{https://github.com/pedrota2000/NNHolo}{NNHolo} \cite{NNHolo}.

\bibliographystyle{JHEP}

\providecommand{\href}[2]{#2}\begingroup\raggedright\endgroup

\end{document}